\documentclass[11pt,a4paper]{JHEP3}

\usepackage[latin1]{inputenc}
\usepackage[english]{babel}
\usepackage{amsmath,amssymb}
\usepackage{graphicx}
\usepackage{cite}
\usepackage{place ins}

\newcommand{\as}{\alpha_s}

\newcommand{\beq}{\begin{equation}}
\newcommand{\eeq}{\end{equation}}
\newcommand{\bea}{\begin{eqnarray}}
\newcommand{\eea}{\end{eqnarray}}
\newcommand{\bdm}{\begin{displaymath}}
\newcommand{\edm}{\end{displaymath}}
\def\as{\alpha_s}

\def\ord{{\cal O}}

\def\d{\partial}

\def \d{{\rm d} }
\def \d0 {D\O \;}

\title{On jet mass distributions in Z+jet and dijet processes at the LHC}

\author{Mrinal Dasgupta, Kamel Khelifa-Kerfa \\Consortium for Fundamental Physics, \\ School of Physics \& Astronomy, University of Manchester, \\ Manchester M13 9PL, United Kingdom\\\email{mrinal.dasgupta@manchester.ac.uk,  kamel.khelifa@hep.manchester.ac.uk}}
\author{Simone Marzani and Michael Spannowsky\\ Institute for Particle Physics Phenomenology, Durham University,\\ Durham DH1 3LE, United Kingdom\\
\email{simone.marzani@durham.ac.uk,  michael.spannowsky@durham.ac.uk}}
\preprint{DCPT/12/96 \\ IPPP/12/48\\ MAN/HEP/2012/08}

\abstract{The mass distribution of jets produced in hard processes at the LHC plays an important role in several jet substructure related studies 
involving both Standard Model and BSM physics, especially in the context of 
boosted heavy particle searches. We compute analytically the jet-mass distribution for both Z+jet and dijet processes, for QCD jets defined in the anti-$k_t$ algorithm with an arbitrary radius $R$, to next-to-leading logarithmic accuracy and match our resummed calculation to 
full leading-order results. We note the important role played by initial state radiation (ISR) and non-global logarithms explicitly computed here for the 
first time for hadron collider observables, as well as the jet radius dependence of these effects. We also compare our results to standard Monte Carlo event generators and discuss directions for further studies and phenomenology.}

\begin{document}

\section{Introduction}
Studies of jet substructure have become an area of great interest and 
much activity in the context of LHC phenomenology \cite{boost1,boost2}. The 
primary reason for this relatively recent explosion of interest has been the 
observation that massive particles at the LHC can emerge with large boosts 
either due to the fact that one has an unprecedentedly large centre-of--mass energy $\sqrt{s}$, which consequently enables access to a large range of particle transverse momenta, or due to the possibility that one may have very heavy new BSM particles that could decay to lighter Standard Model particles which thus emerge highly boosted. Such particles with energies or, equivalently, transverse momenta much greater than their mass $p_T \gg m$, would decay to products that are collimated and a significant fraction of the time would be recombined by 
jet algorithms into single jets. One is then faced with the problem of 
disentangling more interesting signal jets from those that are 
purely QCD background jets. In this context, detailed studies of jet substructure had been proposed some time ago \cite{Seymour} as having the potential to help separate signal jets (for instance, those arising from hadronic Higgs decays) from QCD background. More recent and detailed analyses \cite{BDRS,atlas_HZ,PSS} subsequently revealed the great potential offered by substructure studies in the specific case of boosted Higgs searches and paved the way for other substructure based studies in many different new physics contexts, which continue to emerge.

However, one important issue that is a significant hinderance to the success of studies that try to exploit detailed knowledge of jet substructure is the complexity of the LHC as a physics environment, particularly in the context of strong interactions. For example, while looking for signal mass-peaks in the vicinity of a certain value of mass, $m$, one is inevitably led to consider the distribution of background QCD jets around that value. In fact a precise knowledge of jet-mass distributions and the related role of cuts on jet masses are extremely crucial for a wide variety of substructure studies. At the LHC one faces formidable obstacles as far as an accurate description of such spectra is concerned. Firstly, from the viewpoint of QCD perturbation theory, the most reliable tool at our disposal, one is confronted with large logarithms due to the multi-scale nature of the distribution at hand, which are as singular as $\frac{1}{m_J} \alpha_s^n \ln^{2n-1} \frac{p_T}{m_J}$, i.e double logarithmic, plus less singular terms. For boosted studies involving values $p_T \gg m_J$ such logarithms dominate the perturbative expansion and hence fixed-order tools like \textsc{Nlojet}{\footnotesize ++} \cite{nlojet}, which are typically heavily relied upon for accurate QCD predictions, become at worst invalid or at best of severely limited use. To make matters worse, there are non-perturbative effects such as the underlying event and also the issue of pile-up, which leads to a contamination of jets resulting in an often very significant worsening of any perturbative description. 

In light of the above discussion, it is clear that the tools one can most 
readily use to estimate distributions of quantities such as jet masses are in 
fact Monte Carlo event generators. The parton showers encoded in these event generators derive from first principles of QCD and offer a partial resummation of the large logarithms we mentioned before. They can be combined with fixed-order NLO results \cite{POWHEG,MC@NLO} to yield descriptions that also describe accurately regions of phase space where the logarithms in question may not be entirely dominant. Moreover, they include hadronisation models as well as a model dependent treatment of other effects such as the underlying event where the parameters of the model are extensively tuned to various data sets, to render them more accurate descriptions. While such tools are very general and hence of immense value in addressing the complexity of the LHC environment, the level of precision they offer may be considered a still open question. On the perturbative 
side the accuracy of the logarithmic resummation represented by parton showers is not clear. While leading logarithms (double logarithms in the present case) are understood to be guaranteed, at the level of next-to--leading or 
single logarithms (NLL level) the showers are not expected to provide a complete description. For example it is well known that parton showers work in the leading colour (large $N_c$) approximation as far as large-angle soft radiation is concerned, while the jet mass distributions we discuss here receive single 
logarithmic contributions from such effects which start already at leading 
order in $\alpha_s$ and contain subleading colour terms. As we shall show 
single-logarithmic terms on the whole have a large impact on the final result 
and thus it is a little disconcerting to note that they will not be fully accommodated in current shower descriptions. 

Also the event generators we are considering have different models both for parton showers as well as for non-perturbative effects such as the underlying event and hadronisation and while it is always possible to tune the parameters in each event generator to experimental data, a comparison of the separate physics ingredients of each program often reveal large differences which, in certain cases, does not inspire much confidence in the accuracy of the final description, as far as QCD predictions are concerned. As one example of such differences we can refer the reader to the studies of non-perturbative effects in jets at hadron colliders~\cite{Dassalmag}, where large differences were pointed out between the \textsc{Herwig} and \textsc{Pythia} underlying event estimates at Tevatron energies. A perhaps even more directly relevant example is the recent comparison by the ATLAS collaboration of their data on jet masses and shapes \cite{ATLASjetmass, ATLASjetshape, ATLASboosted} to predictions from \textsc{Herwig}{\footnotesize++} and \textsc{Pythia} which for the jet mass case, for example, do not agree very well with one another, with \textsc{Pythia} describing the data better. An understanding of the origin of such differences is certainly an important issue in order to gain confidence in the use of Monte Carlo tools for future LHC phenomenology. Hence the somewhat black-box nature of event generators is an issue which can makes sole reliance on these tools dangerous in the long run. 

Another avenue that can be explored in terms of theoretical predictions is the possibility for analytical resummation of large logarithms. While the 
techniques needed to achieve such resummation currently apply to a more limited number of observables, while event generators are far more general purpose tools, where possible, resummation alleviates some of the present difficulties inherent in an event generator based approach. The typical accuracy of analytical resummation is usually at least NLL with some results having been obtained even up to the $\mathrm{N^3LL}$ level \cite{SchwaBec}. It is also possible and 
straightforward in principle to match these resummed calculations to NLO estimates so as to have an accurate prediction over a wide range of observable values. Resummation has been a valuable tool in achieving high precision in several QCD studies ranging from pioneering studies of LEP event shapes \cite{CTTW} to current studies involving hadron collider event shapes \cite{BSZpheno}. In this paper we carry out such a resummation for the case of the jet mass in Z+jet and dijet processes at the LHC.

If one takes the study of jet-mass distributions as a particular case of studying shapes of jets produced in multi-jet events (see for instance Refs.~\cite{EHLVW1, EHLVW2}) then it is clear that for substructure studies, resummation of the kind that we perform here for the jet mass, will be an important tool in achieving a precise description of the internal structure of jets. For the same reason, all the issues one encounters in the present study and the solutions we propose here to those problems, will also be of general relevance in the wider context of resummed calculations as substructure tools. There have in fact been several recent attempts to address the issue of resummation for jet masses and other substructure observables such as angularities \cite{EHLVW1,EHLVW2, yuan1,yuan2}. In Refs.~\cite{EHLVW1,EHLVW2} it was proposed to study angularities of one or more jets produced in multijet events but calculations were only carried out for jets produced in $e^{+}e^{-}$ annihilation. The calculations carried out in these references also omitted important contributions at single-logarithmic level, the so called non-global logarithms \cite{DassalNG1,DassalNG2}, as was explained in some detail in Ref.~\cite{BDKM}. Further calculations for jet-mass observable definitions in $e^{+}e^{-}$ annihilation were carried out in Ref.~\cite{KSZ1,KSZ2}. References \cite{yuan1,yuan2} on the other hand attempt to address the issue for the case of hadron collisions but employ calculations that are not complete to next-to--leading logarithmic accuracy, taking only into account the collinear branchings that generate the so called process independent jet function approximation to the resummed result. In this approach one does not treat the soft large-angle effects that arise from initial state radiation (ISR) or address the important issue of non-global logarithms or the dependence on the jet algorithm explained in Ref.~\cite{BDKM}, and hence cannot be considered sufficient for a reasonable phenomenological description of data.  

In our current paper we address both issues of ISR and non-global contributions and demonstrate their significance not just as formal NLL terms but also from the perspective of numerics and the accuracy of the final description for potential comparisons to data. We consider specifically the jet mass distribution of jets produced in two different hard processes: Z+jet production where it will be a background to the case of associated boosted Higgs production, 
with Higgs decay to $b \bar{b}$ and the case of jet production in dijet LHC 
events. We carry out a resummed calculation including the ISR contributions as a power series in jet radius $R$, while the non-global logarithms are calculated exactly at leading order (i.e. order $\alpha_s^2$)  and then resummed in the leading $N_c$ approximation as was the case for DIS single hemisphere event shapes studied phenomenologically in Ref.~\cite{DasSalDIS}. We demonstrate that developing calculations for $e^{+}e^{-}$ variables and carrying them over to the LHC with neglect of process dependent ISR and non-global logarithms can yield large differences with the full resummation which correctly includes these effects. Moreover, our calculations have an advantage also over Monte Carlo event generators in that we retain the full colour structure of the ISR terms resorting to the leading $N_c$ approximation only for the non-global terms starting from order $\alpha_s^3$. The accuracy that we achieve in our resummed exponent should then be comparable to that which yielded a good description of DIS event shape data \cite{DasSalDIS}. We also match our results to leading order QCD predictions so as 
to account for those terms which are not enhanced by large logarithms but may be important at larger values of jet mass, i.e. away from the peak of the distribution. 
The calculations of this paper are valid for jets defined in the anti-$k_t$ algorithm \cite{antikt}. For jets defined in other algorithms such as the $k_t$ \cite{kt1,kt2}, and Cambridge--Aachen algorithms \cite{CAM,CA} 
the role of gluon self-clustering effects greatly complicates the single-logarithmic resummation (see e.g the discussions in Refs.~\cite{BanDas05,BanDasDel,KKK, KWZ1,KWZ2}). For such algorithms analytical resummation to single logarithmic accuracy is still beyond the current state-of--the art and we shall hence not treat them. 

The calculations in the present paper also stop short of achieving the accuracy that was obtained for single-hemisphere DIS event shapes in one aspect. 
While we achieve the same accuracy as the DIS case for the resummed exponent, we do not yet obtain the NNLL accuracy in the expansion of the resummation as is achieved for most global event shapes in $e^{+}e^{-}$ and DIS in the leading $N_c$ limit \cite{Dassalrev} as well as in hadron collisions \cite{BSZpheno}. In other words our current resummation would not guarantee obtaining the $\alpha_s^2 L^2$ terms in the expansion, that arise from a cross-talk between a constant coefficient function $ \alpha_s C_1$, which corrects the resummation off just the Born configuration, and the $\alpha_s L^2$ term of the Sudakov form factors one obtains for jet masses. A proper treatment of such constant coefficients requires further work which we shall carry out in a subsequent article. At that stage we will also be in a position to carry out an NLO matching which, at least from the perturbative viewpoint, will give us an answer that will represent the state of the art for non-global observables. We do however estimate in this paper the possible effect of correcting for the coefficient function on our present predictions for the case of Z+jet production. In order to proceed to full NLL accuracy one would also need to understand non-global logarithms beyond the leading $N_c$ approximation but this is a much longer term goal. In the meantime we believe that the predictions we obtain here and certainly after forthcoming NLO matching will be sufficiently accurate so as to render them valuable for phenomenological studies. For the present moment we compare our resummed results to results from shower Monte Carlos and comment on the interesting features that emerge.

We organise this article as follows: in the following section we outline the 
general framework for our resummed results indicating the separation between 
the global piece and the non-global terms. Following this, we derive in more 
detail the results for the global terms for both the case of Z+jet and dijet production. In section~\ref{sec:nglogs} we detail the results of our calculation for the 
non-global component at fixed-order and at all orders in the large $N_c$ limit. In section~\ref{sec:Zjet} we plot our final results for the case of Z+jet production, having carried out a leading-order matching and commented on the impact of various contributions to the resummed exponent such as ISR and non-global logarithms. We also discuss the expected effect on our results of a proper treatment of the coefficient $C_1$, by treating its contribution in different approximations. Lastly, for the Z+jet case we compare our results to Monte Carlo estimates from a variety of event generators. In section~\ref{sec:dijets} we discuss final results for the case of dijet production with matching to leading-order calculations. Finally in 
section~\ref{sec:conclusions} we arrive at our conclusions. Detailed calculations and explicit resummation formulae are collected in the Appendices.

\section{General framework} \label{sec:framework}
The purpose of this section is to outline the overall structure of the resummed results that we have computed for both jet production in association with a 
vector boson and dijet processes at hadron colliders. The notation we find most convenient to adopt is the one developed and used in Refs.~\cite{BSZcaesar,BSZpheno} for the case of global event shape variables in hadron collisions. The extra ingredient specific to our calculation of jet masses is essentially that 
the observables we address are non-global, so that certain specifics shall of course differ from the case of global event shapes, which we shall highlight, where relevant.

For the case of jet production in association with a Z boson we shall thus 
examine a distribution of the form 
\begin{equation}
\frac{1}{\sigma} \frac{d \sigma}{dm_J^2}
\end{equation}
where $m_J^2$ is the  jet-mass squared of the highest $p_T$ jet recoiling against the Z boson. For dijet production we shall instead adopt a different 
observable definition and study essentially the jet mass distribution averaged over the highest and second highest transverse momentum jets:
\begin{equation}
\frac{1}{\sigma} \left(\frac{d \sigma}{dm_{J1}^2}+\frac{d \sigma}{dm_{J2}^2} \right)
\end{equation}
where $m_{J1}^2$ and $m_{J2}^2$ are the squared masses of the highest and second highest $p_T$ jets, respectively. 

At Born level, for the Z+jet case, we have a single parton recoiling against the Z boson. If we restrict ourselves  to a single-jet final state, then the jet-mass distribution for small jet masses is generated by soft and collinear 
emission off the hard Born configuration whose colour content is provided by the two incoming partons and the final state parton that initiates the jet. 
In contrast, for the case of dijet production soft and collinear emissions 
around the Born configuration, involve an ensemble of four coloured particles, with two incoming partons and two outgoing partons corresponding to the jets. 

While for global observables, such as event shapes, resummation off the hard Born configuration is all that matters, in the present case it is obvious that small jet masses can be produced in events with any jet multiplicity, that represent higher order corrections to the basic Born configurations we address. To restrict oneself to addressing just the Born configuration one can, for example, impose a veto scale $p_{T0}$, as suggested in  Ref.~\cite{EHLVW1}. However, depending on the value of this scale one may then need to also resum the consequent logarithms involving the veto scale (see Refs.~\cite{EHLVW1, EHLVW2} for a discussion). Even if one chooses to adopt this procedure, the calculations we carry out and report in this paper shall still form the basis of the resummed answer, but will need to be modified to account additionally for the imposition of a veto. In the present article, we do not impose a veto but note that any additional production of non-soft, non-collinear particles (e.g the Z+2 jet correction terms to the leading Z+jet process) will be associated with a suppression factor of $\alpha_s(p_T)$ relative to the Born term, for each additional jet, where $p_T$ is the typical transverse momentum of the additional jet. This means that to the accuracy of our present paper (and indeed the accuracy of most current resummed calculations) we shall need to account for only the order $\alpha_s$ correction to the Born term supported by a form factor involving only the double logarithmic (soft {\emph{and}} collinear) component of the jet mass resummation. Thus, we never have to discuss, to our accuracy, the complex issue of soft wide-angle gluon resummation off an ensemble other than the Born configuration. The role of correction terms to the basic Born level resummation shall be discussed in more detail later in the article. 

Next, following the notation of Refs.~\cite{BSZcaesar,BSZpheno} and denoting 
the Born kinematical configuration by ${\mathcal{B}}$ we write, for a fixed Born configuration, the cross-section for the squared jet mass to 
be below some value $v p_T^2 $, as 
\beq \label{sigmadef}
\frac{d\Sigma^{(\delta)}(v)}{d \mathcal{B}}= \int  d m_J^2 \frac{d^2\sigma^{(\delta)}}{d \mathcal{B} d m_J^2 } \Theta(v p_T^2 -m_J^2) . 
\eeq
The label $\delta$ corresponds to the relevant production channel at Born level, i.e the flavour structure of the underlying $2 \to 2$ Born process. We have also introduced the dimensionless variable $v=m_J^2/p_T^2$, with $p_T$ the transverse momentum of the measured jet. One can integrate over the Born configuration 
with a set of kinematical cuts denoted by $\mathcal {{H}} (\mathcal{B})$ to obtain the integrated cross-section 
\begin{equation}
\Sigma^{(\delta)}(v) = \int d\mathcal{B} \frac{d \Sigma^{(\delta)}(v)}{d\mathcal{B}} {\mathcal{H}}(\mathcal{B}),
\end{equation} 
where, as should be clear from the notation, $\frac{d\Sigma^{\left(\delta \right)}}{d \mathcal{B}}$ is the fully differential Born cross-section (i.e the leading order cross-section at fixed Born kinematics) for the subprocess labelled by $\delta$. 
We can then sum over Born channels $\delta$ to obtain $\Sigma(v)$ the integrated jet mass cross-section. However, the above result differs from the case of global observables considered in Refs.~\cite{BSZcaesar, BSZpheno}, in that it is correct only as far as the resummation of logarithms off the Born configuration is concerned and not at the level of constant terms which can arise from the higher jet topologies we mentioned before, that are not related to the Born configuration. When addressing the issue of constant corrections we shall thus need to account in addition to the above, for jet production beyond the Born level. For the present we focus on the basic resummation and hence work only with the Born level production channels as detailed above.

Follwing Ref.~\cite{BSZcaesar} (for $v \ll 1$) we then write
\begin{equation}
\label{eq:fact}
\frac{d\Sigma^{\left(\delta \right)}(v)}{d \mathcal{B}} = \frac{d\sigma_0^{\left(\delta \right)}}{d \mathcal{B}} f_{\mathcal{B}}^{(\delta)}\left(1+\mathcal{O} \left(\alpha_s \right) \right).
\end{equation}
The resummation is included in the function $f_{\mathcal{B}} ^{(\delta)}$ and has the usual form \cite{CTTW}:
\begin{equation}
f_{\mathcal{B}}^{(\delta)} = \exp \left [Lg_1(\alpha_s L)+g_2(\alpha_s L)+\alpha_s g_3(\alpha_s L) +\cdots \right]
\end{equation} 
where $g_1$, $g_2$ and $g_3$ are leading, next-to--leading and next-to--next to leading logarithmic functions with further subleading terms indicated by the 
ellipses and $L=\ln \frac{1}{v}$. 

For the observable we study here, namely non-global jet mass distributions, 
the function $g_1$ is generated simply by the time-like soft and collinear branching of an outgoing parton and depends only on the colour Casimir operator of the parton initiating the jet, while being independent of the rest of the event. The function $g_2$ is much more complicated. It has a piece of  pure hard-collinear origin, which, like the leading logarithmic function $g_1$, only depends on the colour charge of the parton initiating the jet and factorises from the rest of the event. In the collinear approximation, combining the soft-collinear terms of $g_1$ and the hard-collinear terms included in $g_2$ we recover essentially the jet functions first computed for quark jets in $e^{+}e^{-}$ annihilation in \cite{CTTW}. However, for complete single logarithmic accuracy one has to consider also the role of soft wide-angle radiation. The function $g_2$ receives a pure soft large-angle contribution also due to emissions from hard partons other than the one initiating the jet. For the Z+jet case this piece would be generated by coherent soft wide-angle emission from a three hard parton ensemble, consisting of the incoming partons and the outgoing hard parton (jet). For the case of dijet production, we have instead to consider an ensemble of four hard partons and the consequent soft wide-angle radiation has a non-trivial colour matrix structure~\cite{KOS}, as for global hadronic dijet event shapes. 

Other than the above effects, which are all present for global event shapes and which are all generated by a single soft emission, the function $g_2$ receives another kind of soft contribution, starting from the level of two soft gluons. 
Since we are looking in the interior of a jet rather than the whole of phase 
space, our observable is sensitive to soft gluons outside the jet region 
emitting a much softer gluon into the jet. While for a global observable such a much softer emission would cancel against virtual corrections, in the present case it makes an essential contribution  to the jet mass, triggering single logarithms in the jet mass distribution. These single logarithms (non-global logarithms) cannot be resummed by traditional methods which are based essentially on single gluon exponentiation. In fact a resummation of non-global terms, valid in the large $N_c$ limit, which corresponds to solving a non-linear evolution equation~\cite{BMS}, can be obtained by means of a dipole evolution code \cite{DassalNG1}. We carry out such a resummation in this article but do not attempt to 
address the issue of the subleading $N_c$ corrections which are as yet an 
unsolved problem. Since non-global logarithms are next-to--leading and we are in fact able to obtain the full colour structure for them up to order $\alpha_s^2$, it is only single logarithmic terms starting at order $\alpha_s^3$ where one needs to use the large $N_c$ approximation. One may thus expect that for 
phenomenological purposes an adequate description of non-global effects will be provided by our treatment here, as was the case for DIS event shape variables studied in Ref.~\cite{DasSalDIS}.

In the next section we shall generate the entire result, except for the non-global terms, which we shall correct for in a subsequent section. The results of the 
next section correspond to the answer that would be obtained if the observable were a global observable and include process independent soft and 
hard-collinear terms alluded to above, as well as a process dependent soft wide-angle piece, which also depends on the jet radius $R$. This soft wide-angle piece, which starts at order $\alpha_s$, is calculated with full colour structure whereas one would expect that in Monte Carlo event generators only the leading $N_c$ terms are retained, thus implying higher accuracy of the results we obtain here.

\section{The eikonal approximation and resummation} \label{sec:eikonal}
In the current section we shall consider the emission of a soft gluon by an ensemble of hard partons in the eikonal approximation. In this limit one can consider the radiation pattern to be a sum over dipole emission terms \cite{QCDESW}. Our strategy is to calculate the individual dipole contributions to the jet mass distribution and then to sum over dipoles to obtain results for both the 
Z+jet case as well as the dijet case. While the sum over dipoles shall 
generate both the soft-collinear and soft wide-angle terms we mentioned in the preceding section, we shall need to extend our answer 
to include also the relevant hard-collinear terms. Once this is done, the only remaining source of single logarithmic terms will be the non-global contribution to the single-logarithmic function $g_2$, which we shall address in detail in the following section.

We shall consider the most general situation that we need to address with all dipoles formed by two incoming partons and two outgoing hard partons. Clearly,
for the Z+jet case one would have only a single hard parton in the final state, with the other parton replaced by a massive vector boson and hence for this case we will exclude the dipole contributions involving the recoiling jet.

The squared matrix element for emission of a soft gluon $k$ by a system of hard dipoles is described, in the eikonal approximation, as a sum over contributions from all possible colour dipoles:
\beq
 \label{eq:dipsum}
 \left| {\mathcal{M}}_{\delta}\right|^2 = \left| {\cal M}_{\mathcal{B},\delta} \right|^2 
 \sum_{(ij )\in \delta } C_{ij} \, W_{ij}(k)~,
\eeq 
where the sum runs over all distinct pairs $(ij)$ of hard partons present in the flavour configuration $\delta$, or equivalently, as stated before, over all dipoles. The quantity $\left| {\cal M}_{\mathcal{B,\delta}}\right|^2$ is the squared matrix element for the Born level hard scattering, which in our case has to be computed for each separate partonic subprocess $\delta$ contributing to the jet distribution and contains also the dependence on parton distribution functions. The contribution of each dipole $W_{ij}$ is weighted by the colour factor $C_{i j} $, which we shall specify later, while the kinematic factor $W_{ij} (k)$ is explicitly given by the classical antenna function
\beq
 W_{ij} (k) = \frac{\alpha_s \left( \kappa_{t, i j} \right)}{2 \pi}
 \frac{p_i \cdot p_j}{(p_i \cdot k)(p_j \cdot k)}.
 \label{eikant}
\eeq
In the above equation $\alpha_s$ is defined in the bremsstrahlung scheme \cite{Catani:1990rr}, and its argument is the invariant quantity $\kappa_{t, i j}^2 = 2 (p_i \cdot k)(p_j
\cdot k)/(p_i \cdot p_j)$, which is just the transverse momentum with respect
to  the dipole axis, in the dipole rest frame.
We note that in the eikonal approximation, as is well known, the Born level production of hard partons in the relevant subprocess $\delta$, factorises from the production of soft gluons described by the antenna functions $W$. The squared matrix element $|M_{\mathcal{B,\delta}}|^2$ essentially produces the quantity 
$\frac{d\Sigma_0^{\left(\delta \right)}}{d \mathcal{B}}$ while the $W$ functions 
start to build up the exponential resummation factor $f_{\mathcal{B}}^{(\delta)}$ referred to in Eq.~\eqref{eq:fact}. In what follows below we shall focus on the various components of the resummation in more detail and in particular carry out the calculations from the individual dipole terms. 

\subsection{Exponentiation: the Z+jet case}
We have mentioned above the antenna structure of soft gluon emission from a system of hard emitting dipoles. It is well understood by now that if one ignores configurations corresponding to non-global logarithms (in other words those that stem from emission regions where gluons are ordered in energy but not in angle), then to single-logarithmic accuracy there is an exponentiation of the 
one-gluon emission terms described in the preceding section, as well as the corresponding virtual corrections that have the same colour and dynamical 
structure but contribute with an opposite sign so as to cancel the divergences of real emission.  We note that for the case of Z+jet we are dealing with a hard parton ensemble with three partons, two incoming and one corresponding to the triggered jet. In this case the colour factors $C_{i j} = -2\left({\bf T}_i . {\bf T}_j \right)$ (with the ${\bf T}_i$ being SU(3) generators), that accompany the dipole contributions, can be straightforwardly expressed in terms of quark and gluon colour charges. Taking account of virtual corrections and the role of multiple emissions generating the jet mass one can write the result in a form that is familiar from the earliest studies of jet masses in $e^{+}e^{-}$ annihilation \cite{CTTW}
\begin{equation} \label{fglobalZ}
f_{\mathcal{B}, { \rm global}}^{(\delta )} = \frac{\exp[-\mathcal{R}_\delta-\gamma_E \mathcal{R}_\delta']}{\Gamma(1+\mathcal{R}_\delta')},
\end{equation}
where the subscript above denotes that we are considering the global term only i.e ignoring all non-global corrections to $f_{\mathcal{B}}^{(\delta)}$.

The function $(-\mathcal{R})$ represents the exponentiation of the single-gluon contribution after cancellation of real-virtual divergences, while $\mathcal{R}'$ is the logarithmic derivative of $\mathcal{R}$, $\partial_L {\mathcal{R}}$ to be evaluated to our accuracy simply by accounting for the leading logarithmic terms in $\mathcal{R}$. The terms involving $\mathcal{R}'$ arise due to the fact that direct exponentiation only occurs for the Mellin conjugate of the variable $v$. To single-logarithmic accuracy one can invert the Mellin transform analytically by performing a Taylor expansion of the Mellin space result to first order and integrating over the Mellin variable, resulting in the form written above \cite{CTTW}.

One has for $\mathcal{R}_\delta$ the result:
\begin{equation}
\mathcal{R}_\delta = \sum_{(ij)\in \delta}\int C_{ij} \, {dk_t} k_t \, d\eta \, \frac{d\phi}{2\pi} \, W_{ij}(k) \, \Theta \left(v(k)-v \right),
\end{equation} 
where we have introduced the integral over the momentum of the emitted gluon $k$ and the step function accounts for the fact that real-virtual cancellations  
occur below a value $v$ of the normalised squared jet mass, while uncancelled virtual corrections remain above $v$. The function $v(k)$ is just the dependence of jet-mass on the emission $k$. Letting the hard initiating parton have rapidity $y$ and transverse momentum $p_t$ and denoting by $k_t$, $\eta$ and $\phi$ the transverse 
momentum rapidity and azimuth  of the soft gluon $k$ we have (when the hard parton and gluon are recombined to form a massive jet)
\begin{equation} \label{eikonalv}
v(k) = \frac{m_J^2}{|\underline{p}_t+\underline{k}_t|^2} = \frac{2 k_t}{p_t} \left [ \cosh \left ( 
\eta- y \right )- \cos \phi \right ]+\mathcal{O} \left ( \frac{k_t^2}{p_t^2} \right)
\end{equation}
where we neglect terms quadratic in the small quantity $k_t$.

It now remains to carry out the dipole calculations for the Z+jet case. We have a hard process with two coloured fermions and a gluon or a three-hard--particle antenna, irrespective of the Born channel $\delta$. Let us call $\delta_1$ the Born subprocess with an incoming quark (or anti-quark) and an incoming gluon, which results in a final state coloured quark or antiquark recoiling against the Z boson. Labelling the incoming partons as $1$ (fermion) and $2$ (gluon) and the measured jet as $3$ we have the following colour factors:
\begin{equation}
\label{eq:colfac}
C_{12} = N_c, \; C_{23} = N_c, \; C_{13} = -\frac{1}{N_c}.
\end{equation} 
For the remaining Born channel with an incoming $q \bar{q}$ pair we obtain the same set of colour factors as above but with an interchange of $2$ and $3$, to correspond to the fact that is always the quark-(anti)quark dipole which is colour suppressed.

The calculation of individual dipole terms is carried out in 
Appendix~\ref{app:global}. We use the results obtained there to construct the final answer. Let us focus on the Born channel $\delta_1$ corresponding to an incoming gluon and quark with a 
measured quark jet. In order to combine the various dipoles that contribute to the resummed exponent, we combine the pieces $\mathcal{R}_{ij}$ computed in the appendix weighting them appropriately by colour factors:
\begin{equation}
\mathcal{R}_{\delta_1} = C_{12} \, \mathcal{R}_{12}+C_{13} \, \left( \mathcal{R}_{13}^{\mathrm{soft}} +\mathcal{R}_{13}^{\mathrm{coll.}} 
\right)+C_{23} \, \left ( \mathcal{R}_{23}^{\mathrm{soft}} +\mathcal{R}_{23}^{\mathrm{coll}} \right).
\end{equation}

Using the appropriate dipole results generated by using the eikonal approximation, from Appendix~\ref{app:global} and using the colour factors mentioned in Eq.~\eqref{eq:colfac} we obtain 
\begin{equation}
\mathcal{R}_{\delta_1} = \frac{N_c^2-1}{N_c} \mathcal{R}^{\mathrm{coll.}} + N_c \mathcal{R}^{\mathrm{soft}}_{12}+\frac{N_c^2-1}{N_c}\mathcal{R}^{\mathrm{soft}},
\end{equation}
where we used the fact that $\mathcal{R}_{13}^{\mathrm{coll.}}=\mathcal{R}_{23}^{\mathrm{coll.}}= \mathcal{R}^{\mathrm{coll.}}$ and $\mathcal{R}_{13}^{\mathrm{soft}}=\mathcal{R}_{23}^{\mathrm{soft}}= \mathcal{R}^{\mathrm{soft}}$. Writing the result in terms of the colour factors $C_F$ and $C_A$ and the explicit results for the various dipoles results in the following simple form:
\begin{multline} \label{radiatorCF}
\mathcal{R}_{\delta_1}(v)=2C_F \int \frac{\alpha_s \left( k_{t,J} \right)}{2\pi} \frac{d k_{t,J}^2}{k_{t,J}^2} \ln \left (\frac{R p_te^{-3/2}}{k_{t,J}}   \right) \Theta \left (\frac{k_{t,J}^2}{p_t^2} -v \right) \Theta \left(R^2 -\frac{k_{t,J}^2}{p_t^2} \right)  \\ +2 C_F \int \frac{\alpha_s \left( k_{t,J}\right)}{2\pi} \frac{d k_{t,J}^2}{k_{t,J}^2} \ln \left (\frac{R k_{t,J}}{v p_{t}}   \right) \Theta \left (v-\frac{k_{t,J}^2}{p_t^2} \right ) \Theta \left(\frac{k_{t,J}^2}{p_t^2} -\frac{v^2}{R^2}\right) \\ 
+{R^2} \left(C_A+\frac{C_F}{2} \right) \int_v^1\frac{dx}{x} \frac{\alpha_s(x p_t)}{2 \pi}+\frac{R^4}{144} C_F \int_v^1\frac{dx}{x} \frac{\alpha_s(x p_t)}{2 \pi},
\end{multline}
where $k_{t,J}$ is the transverse momentum of the emitted gluon with respect to the jet.

The above result represents the decomposition of the resummed exponent into a collinear piece contained in the first two lines of the above equation and a soft wide-angle piece. We have included in the collinear piece a term $e^{-3/2}$ in the argument of the logarithm, which corrects the eikonal approximation for hard collinear splittings of the final state quark jet. This correction term emerges from replacing the IR singular (pole part) of the $q \to qg$ splitting function, treated by the eikonal approximation, by the full splitting function. As one would expect this collinear piece, which also contains the leading double logarithms, involves only the colour charge $C_F$ of the parton that initiates the measured massive jet, in this case a quark. The remaining part of the result above is a process dependent soft large angle piece that has a power series expansion in jet radius $R$, which we truncated at the $R^4$ term. We note that the $R^4$ term emerges with a numerically small coefficient and shall make a negligible impact on our final results which can thus be essentially obtained by considering the $\ord(R^2)$ corrections alone. We also note that the calculation of the soft large-angle component of the result should mean that our results are more accurate than those obtained from MC event generators, which would only treat such pieces in a leading $N_c$ approximation.

The case of the other subprocess, which generates a gluon jet in the final state, is totally analogous. The result for the resummed exponent is
\begin{multline} \label{radiatorCA}
\mathcal{R}_{\delta_2}(v)=2C_A \int \frac{\alpha_s \left( k_{t,J} \right)}{2\pi} \frac{d k_{t,J}^2}{k_{t,J}^2} \ln \left (\frac{R p_te^{-2 \pi \beta_0/C_A}}{k_{t,J}}   \right) \Theta \left (\frac{k_{t,J}^2}{p_t^2} -v \right) \Theta \left(R^2 -\frac{k_{t,J}^2}{p_t^2} \right)  \\ +2 C_A \int \frac{\alpha_s \left( k_{t,J}\right)}{2\pi} \frac{d k_{t,J}^2}{k_{t,J}^2} \ln \left (\frac{R k_{t,J}}{v p_{t}}   \right) \Theta \left (v-\frac{k_{t,J}^2}{p_t^2} \right ) \Theta \left(\frac{k_{t,J}^2}{p_t^2} -\frac{v^2}{R^2}\right) \\ 
+{R^2} \left(2 C_F-\frac{C_A}{2} \right) \int_v^1\frac{dx}{x} \frac{\alpha_s(x p_t)}{2 \pi}+\ord(R^4).
\end{multline}
In order to achieve NLL accuracy, the remaining integrals in Eq.~(\ref{radiatorCF}) and Eq.~(\ref{radiatorCA}) must be performed with the two-loop expression for the running coupling. We obtain:
\begin{eqnarray} \label{Zjet_rad}
\mathcal{R}_{\delta_1}&=& -C_F \left (L f_1+f_2 + f_{{\rm coll},q} \right) - R^2 f_{\rm l.a.} \left (C_A+\frac{C_F}{2} \right) + \ord(R^4), \nonumber \\
\mathcal{R}_{\delta_2}&=& -C_A \left (L f_1+f_2 + f_{{\rm coll},g} \right) -R^2 f_{\rm l.a.} \left ( 2 C_F -\frac{C_A}{2}\right) + \ord(R^4), \nonumber \\
\end{eqnarray}
with $L=\ln \frac{R^2}{v}$. Explicit results for the functions $f_i$ are collected in Appendix~\ref{app:resum}.

\subsection{Exponentiation: the dijet case}
We now turn our attention to the process
\begin{equation}
p(P_1) + p(P_2) \to J (p_3)+J(p_4) + X,
\end{equation}
where we want to measure the mass of the two leading jets. For convenience, we fix the kinematics of the two (back-to-back, in the eikonal limit) leading jets, i.e. their transverse momentum $p_T$ and their rapidity separation $|\Delta y|$.
The calculation of the dipoles in the eikonal limit proceeds in the same way as in the  Z+jet case that we have previously analysed. The main difference is 
the more complicated colour algebra, that leads to a matrix-structure of the resummed result. The formalism to perform the resummation in the presence of more than three hard partons was developed in~\cite{KOS}. For each partonic sub-process we need to fix a colour basis and find the corresponding representations of the colour factors ${\bf T}_i . {\bf T}_j$. For each partonic subprocess, the resummed exponent takes the form
\begin{equation} \label{dijets_rc}
f_{\mathcal{B}, { \rm global}}^{(\delta )} = \frac{1}{ {\rm tr} \, H_{\delta} } \sum_{J=3,4} {\rm tr} \left[ \frac{ H_{\delta} e^{-\left( \mathcal{G}_{\delta,J}+\gamma_E \mathcal{G}'_{\delta,J}\right)^{\dagger}}S_{\delta,J} e^{-\mathcal{G}_{\delta,J}-\gamma_E \mathcal{G}'_{\delta,J}}+(\Delta y \leftrightarrow - \Delta y)  }{\Gamma\left(1+2 \mathcal{G}'_{\delta,J} \right)}  \right].
\end{equation}
The matrices $H_\delta$ correspond to the different Born subprocesses and ${\rm tr } H=\frac{d \sigma_0^{(\delta)}}{ d\mathcal{B}}$. We note that the resummed expression in Eq.~(\ref{dijets_rc}) is written in terms of exponentials that describe the colour evolution of the amplitude~\footnote{In the literature this resummed expression is written in terms of an anomalous dimension $\Gamma$, where $\mathcal{G}= \Gamma \xi$ and $\xi$ is the appropriate evolution variable.}, rather than of the cross-section as in the Z+jet case. We obtain
\begin{eqnarray} \label{dijets_rad}
\mathcal{G}_{\delta,3}&=& -\frac{{\bf T}_3^2}{2}\left (L f_1+f_2 + f_{{\rm coll},3} \right) + {\bf T}_1 . {\bf T}_2 f_{\rm l.a.}(2 \pi i +R^2)
\nonumber \\ &&+R^2f_{\rm l.a.}  \left( \frac{1}{4} {\bf T}_3.{\bf T}_4 \tanh^2 \frac{\Delta y}{2}+   \frac{1}{4} ({\bf T}_1.{\bf T}_3+ {\bf T}_2.{\bf T}_3) \right.  \nonumber \\ &&+\frac{1}{2} {\bf T}_1.{\bf T}_4 \frac{e^{\Delta y}}{1+\cosh \Delta y}+ \left. \frac{1}{2} {\bf T}_2.{\bf T}_4 \frac{e^{-\Delta y}}{1+\cosh \Delta y} \right) \nonumber +\ord(R^4),  \nonumber \\
\mathcal{G}_{\delta,4}&=& -\frac{{\bf T}_4^2}{2}\left (L f_1+f_2 + f_{{\rm coll},4} \right) + {\bf T}_1 . {\bf T}_2 f_{\rm l.a.}(2 \pi i +R^2)
\nonumber \\ &&+R^2f_{\rm l.a.}  \left( \frac{1}{4} {\bf T}_3.{\bf T}_4 \tanh^2 \frac{\Delta y}{2}+   \frac{1}{4} ({\bf T}_1.{\bf T}_4+ {\bf T}_2.{\bf T}_4) \right.  \nonumber \\ &&+\frac{1}{2} {\bf T}_2.{\bf T}_3 \frac{e^{\Delta y}}{1+\cosh \Delta y}+ \left. \frac{1}{2} {\bf T}_1.{\bf T}_3 \frac{e^{-\Delta y}}{1+\cosh \Delta y} \right) \nonumber +\ord(R^4). \nonumber \\
 \end{eqnarray}
where the functions $f_i$ are reported in Appendix~\ref{app:resum} and, as before, $L=\ln \frac{R^2}{v}$, $\mathcal{G}'=\partial_L \mathcal{G}$. The collinear part of the result is diagonal in colour space, with a coefficient which is the Casimir of the jet. Large-angle radiation is instead characterised by a more complicated colour structure. We also note the presence of the imaginary phase due to Coulomb gluon exchange. We choose to work in the set of orthonormal bases specified in~\cite{FKM}, to which we remind for the explicit expressions. As a result, all the colour matrices are symmetric and the soft matrix appearing in Eq.~(\ref{dijets_rc}) is the identity $S_{\delta,J}=1$.

As an example, we report explicit results for the scattering of quarks with different flavours $q(i) q'(j) \to q(k) q'(l)$. We work in a normalised singlet-octet basis:
\begin{eqnarray} \label{qqqqbasis}
c_1 &= & \frac{1}{N_c}\delta_{ik} \delta_{jl}\,, \nonumber \\
c_2 &=& \frac{1}{\sqrt{N_c^2-1}} \left(\delta_{il} \delta_{jk}-\frac{1}{N_c}\delta_{ik} \delta_{jl} \right).
\end{eqnarray}
In the $t$ channel ($\Delta y>0$), the hard scattering matrix is given by
\begin{equation} \label{qqpqqpH}  
H(t,u)= \frac{4}{N_c^2}\left(\begin{array}{cc}
0 & 0\\
  0 &  \frac{u^2+s^2}{t^2}
 \end{array}
   \right)\,.
\end{equation}
We have that ${\bf T}_3^2={\bf T}_4^2=C_F$ and the other colour matrices are
\begin{eqnarray}
{\bf T}_1 . {\bf T}_2= {\bf T}_3. {\bf T}_4&=& \left(\begin{array}{cc}
0 & \frac{\sqrt{N_c^2-}1}{2 N_c}\\
 \frac{\sqrt{N_c^2-1}}{2 N_c} &-\frac{1}{N_c}
 \end{array}   \right)\,, \nonumber \\
 {\bf T}_1 . {\bf T}_3= {\bf T}_2. {\bf T}_4&=& \left(\begin{array}{cc}
-C_F & 0\\
 0 &\frac{1}{2N_c}
 \end{array}   \right)\,, \nonumber \\
 {\bf T}_2 . {\bf T}_3= {\bf T}_1. {\bf T}_4&=& \left(\begin{array}{cc}
0 & -\frac{\sqrt{N_c^2-}1}{2 N_c}\\
 -\frac{\sqrt{N_c^2-1}}{2 N_c} &\frac{1}{2N_c}-C_F
 \end{array}   \right)\,.
 \end{eqnarray}

\subsection{The constant term $C_1$} \label{sec:C1}
In order to achieve the NNLL accuracy in the perturbative expansion, that is common for most resummation studies of event shape variables, one must consider also the $\mathcal{O}(\alpha_s)$ corrections, which are not logarithmically enhanced in the small jet-mass limit and their cross-talk with double logarithmic terms arising from the Sudakov form factors. The constant terms can be expressed as :
\bea \label{C1def}
\alpha_s \mathcal{C}_{1}^{(\delta)}&=& \lim_{v \to 0} \left[ \Sigma_{\rm NLO}^{(\delta)}(v)-\Sigma^{(\delta)}_{{\rm NLL},\as}(v)\right]=
 \lim_{v \to 0} \left[ \int_0^{v}\frac{d \sigma^{(\delta)}}{d v}d v-\Sigma^{(\delta)}_{{\rm NLL},\as}(v)\right] \nonumber \\ &=&
 \lim_{v \to 0} \left[ \sigma^{(\delta)}_{\rm NLO}-\int_{v}^{v_{\rm max}}\frac{d \sigma^{(\delta)}}{d v}d v-\Sigma^{(\delta)}_{{\rm NLL},\as}(v)\right] \nonumber \\ &=&
\sigma^{(\delta)}_{\rm NLO}+\lim_{v \to 0} \left[\int^{v}_{v_{\rm max}}\frac{d \sigma^{(\delta)}}{d v} d v-\Sigma^{(\delta)}_{{\rm NLL},\as}(v)\right].
\eea
If $\delta$ is a partonic channel that is also present at Born level, then we can recover the usual definition of the constant term:
\beq
{C}_{1}^{(\delta)}=\frac{\mathcal{C}_{1}^{(\delta)}}{\sigma_0^{(\delta)}}.
\eeq

The general kinematic, colour and flavour structure of $\mathcal{C}_{1}^{(\delta)}$ can be rather complicated. However, as discussed for global event-shapes in Ref.~\cite{BSZcaesar, BSZpheno}, one can simply multiply together this constant and the appropriate resummed exponent $f_{\mathcal{B}}^{(\delta)}$ previously discussed, essentially because the only relevant terms at NLL originate by the product of $C_{1}^{(\delta)}$  times the double logarithms (soft and collinear) in the exponent, which do not depend on the colour flow in the hard scattering nor on the parton distribution functions. This is also true in the case of the jet-mass we are considering in this paper, with the further complication that we need to specify the flavour of the jet we are measuring, because quark and gluon jets will receive different suppression factors. In particular, new channels open up at relative order $\alpha_s$ which are not related to the Born channels and hence obtaining the contribution of the constant terms separately for these channels is an involved exercise. While we leave the complete determination of $C_1$ in different experimental set-ups, as well as an analysis of the dijet case, for future work, for the case of  the jet mass in Z+jet, where we 
measure the mass of the hardest jet, we shall provide later an estimate of the contribution of the constant terms $C_1$ to the resummed distribution.

\section{Non-global logarithms} \label{sec:nglogs}
\subsection{Fixed order}
As we emphasised before, the jet mass that we study here, is a non-global observable which means that the results presented in the previous section are not sufficient to obtain the next-to--leading logarithmic accuracy that is common for event shape variables in $e^{+}e^{-}$ annihilation. One is 
also required to correct the results for the effect of soft wide-angle emissions arising from gluon branching; in other words correlated gluon emission as opposed to independent emission of soft gluons by the hard particle ensemble. Calculations involving correlated gluon emission have been carried out at fixed-order and all-orders (in the leading $N_c$ limit), in the simpler cases of non-global event shapes in $e^{+}e^{-}$ annihilation (such as the light jet-mass) and DIS. Till date a full calculation even at fixed-order has not been carried out for hadron collisions, which we shall do below, in the context of the jet-mass distribution. We also note that in our previous work we carried out a calculation of non-global logarithms for jet masses of jets produced in $e^{+}e^{-}$ annihilation, in the limit of small jet radius $R \ll 1$, which we argued could serve as a model for the calculation in the hadron collision case. However, in the current work we shall lift the approximation of small $R$ meaning that our calculations should be  useful for jets of any radius. 

To address the correlated emission term at leading-order we need to consider 
the case where one has two energy-ordered soft gluons $k_1$ and $k_2$ such that, for instance,  $\omega_1 \gg \omega_2$, where $\omega_1$ and $\omega_2$ are the respective energies. In previous work involving $e^{+}e^{-}$ annihilation we have addressed this situation by using the fact that the emission of such a two-gluon system off a hard $q\bar{q}$ dipole is described by an independent 
emission term with colour factor $C_F^2$ and a correlated emission term with 
colour factor $C_F C_A$. However, in the present case of multiple hard partons, the emission of a two-gluon system is in principle more involved, since there are several emitting dipoles. In practice, the structure of two-parton emission for the cases we study in this paper (i.e up to $n=4$ hard legs) is known to be remarkably simple~\cite{CatGraz2gluon,BMDZ3jet}.

As an illustrative example we consider again the Z+jet case, with three hard legs and the Born channel $\delta_1$, with an outgoing quark jet. 
The squared matrix element once again contains an independent emission piece, which contributes to the exponentiation of the single gluon result, described by our function $\mathcal{R}$. This leaves the correlated parton emission term which has the structure
\begin{equation}
W^{\mathrm{corr.}}(k_1,k_2) = \frac{N_c}{2} w_{12}^{\left(2\right)}+\frac{N_c}{2} w_{23}^{\left(2\right)}-\frac{1}{2N_c} w_{13}^{\left(2 \right)},
\end{equation}
where $w_{ij}^{\left(2 \right)}$ is the correlated two-gluon emission by a dipole $ij$, and which is the same as for the $q \bar{q}$ case studied in $e^{+}e^{-}$ annihilation. Hence the dipole emission and associated colour structure for a correlated two-parton system is precisely the same as for a single gluon emission \cite{BMDZ3jet}. 

Next we note that, as described for example in Ref.~\cite{BMDZ3jet}, a piece of the correlated two-parton emission contribution actually goes to build up 
the running coupling we have already considered in the exponentiated single gluon contribution $\mathcal{R}$. This piece comprises of gluon splittings into 
equally hard gluons or into a $q \bar q$ pair, which together produce the leading term of the QCD $\beta$ function \footnote{For a non-global observable such hard splittings will also give rise to non-global logarithms below the single logarithmic level, which are subleading from our point of view~\cite{KSZ2,KKK}.}. This leaves us to consider only the soft part of the correlated emission $\mathcal{S}$~\cite{milan}, which describes the production of an energy-ordered two-gluon system. For a global 
observable as is well known, this term produces no effect at single-logarithmic accuracy whereas in the present case it provides us with the first term of the non-global contribution. For a general dipole $ij$ we can explicitly write
\begin{equation}
\label{eq:coll}
\mathcal{S}_{ij} = 2 C_A {\bf T}_i.{\bf T}_j \left (A_{ab}+A_{ba}-A_a A_b \right),
\end{equation}
where
\begin{equation}
\label{eq:ngdip}
A_{ab} = \frac{(ij)}{(i k_a)(k_a k_b)(k_b j)}\,, \quad A_{a} = \frac{(ij)}{(i k_a) (j k_a)}.
\end{equation} 
We note that this term is free of collinear singularities along the hard legs $a$ and $b$ due to cancellation between the various terms of Eq.~\eqref{eq:coll}. The remaining collinear singularity from the $1/\left( k_1.k_2 \right)$ , involving the soft gluons, shall turn out to be integrable for the non-global configurations considered below. Thus to obtain the leading non-global term for the jet mass under consideration, it suffices to carry out dipole calculations using the Eqs.~\eqref{eq:coll} and \eqref{eq:ngdip}, for each of the hard emitting dipoles, just as we did for the single gluon emission case. The results we 
obtain are mentioned and commented on below with the details on the calculation in Appendix~\ref{app:nonglobal}. We report below the coefficients $I_{ij}$ of the non-global single logarithms:
\beq
\int [d k_1] [d k_2]  \mathcal{S}_{ij} \Theta_{k_1 \notin J } \Theta_{k_2 \in J } \equiv C_A {\bf T}_i.{\bf T}_j I_{ij}  \ln^2 \frac{1}{v}.
\eeq

\begin{itemize}

\item {\bf{dipoles involving the measured jet}}

We first consider dipoles involving the measure jet:
\begin{equation}
I_{13}=I_{23}\simeq I_{34} = \frac{\pi^2}{3}+ \alpha_2 R^2+\alpha_4 R^4+\ord(R^6).
\end{equation}
We find that the corrections to the small-$R$ approximation~\cite{BDKM} are numerically small, with $\alpha_2\approx 0$ and $\alpha_4\approx 0.013$. 
The dipole $I_{34}$, which is relevant for the dijet calculation, depends, in principle, on the rapidity separation $\Delta y$ between the leading jets. 
This dependence will be associated with powers of $R$ that appear to make a negligible contribution. 

We note that the result $\frac{\pi^2}{3}$ is exact for $e^{+}e^{-}$ collisions, where one defines the jet in terms of a polar angle. This result is precisely the same as was obtained for hemisphere jet masses in $e^{+}e^{-}$ annihilation and implies that the result obtained does not depend on the position of the jet boundary, a feature that has also been observed in~\cite{HLWZinout}.

%

 \item {\bf{in-in dipole}}
 
Here we are considering a situation where the emitting dipole legs are away 
from the interior of the jet region into which the softest gluon $k_2$ is emitted. This situation is reminiscent of the much studied case of non-global logarithms in energy flow into gaps between jets \cite{DassalNG2,BMS}. In Ref.~\cite{DassalNG2} for instance, energy flow into an inter-jet region (i.e a region between 
but away from the hard legs of an emitting dipole) was considered for specific choices of the geometry of this region, such as a rapidity slice and a patch in rapidity and azimuth. In the present case of course we are studying the same problem but the region being considered here is a circle $(\eta-y)^2+\phi^2$, centred on the measured jet. The result we obtain for the non-global contribution due to the in-in dipole was computed numerically but has the small $R$ behaviour

\begin{equation}
I_{12} \approx 4 \left[1.17 R^2-R^2 \ln 2R +\mathcal{O} \left(R^4 \right)\right ].
\end{equation}
We note that the $R^2$ behaviour arises simply from integrating the emitted softest gluon $k_2$ over the interior of the jet region while the $\ln R$ behaviour is a reflection of the collinear singularity we mentioned above, between the two soft gluons $k_1$ and $k_2$. In practice the above small $R$ approximation is not a good approximation to the full result obtained numerically, for larger values of $R \sim 1$, hence eventually we use the numerical answer rather than the form above.

\item {\bf{in-recoil dipole}}

Here again the situation is similar to the case of the interjet energy flow with the only difference from the previous dipole being the geometry of the hard dipole legs, which are now formed by an incoming parton and an outgoing parton at a finite rapidity with respect tothe beam.
 The result reads
\begin{equation}
I_{12} \approx\frac{\left(1+e^{\Delta y}\right)^2}{\left(1+\cosh \Delta y \right)^2} \left(1.17 R^2 -R^2 \ln 2 R\right)+\frac{(1+e^{\Delta y})}{1+\cosh \Delta y} \kappa (\Delta y) R^2.
\end{equation}
where $\kappa$ depends on rapidity difference $\Delta y =y-y_r$ and is evaluated numerically. In the limit $\Delta y \to \infty$, $\kappa(\Delta y) \to 0$ and one recovers the previous case.
The dipole $I_{24}$ is easily obtained as 
\begin{equation}
I_{24}= I_{14}(-\Delta y).
\end{equation}

\end{itemize}

\subsection{All-order treatment}
In order to perform a complete (NLL) resummation of non-global logarithms, one would need to consider the emission of a soft gluon from an ensemble of any number of gluons. This problem can be treated only in the large-$N_c$ limit~\cite{DassalNG1,BMS}\footnote{An alternative approach that can be found in the literature consists in an expansion in the number of out-of-jet (in this case), gluons, keeping the full colour structure, see for instance~\cite{FKS1,FKS2,FKM,DFMS}.}. For this study, we have adapted the dipole-evolution code used in~\cite{DassalNG1} to perform the resummation in the case of jet-masses. 

The code developed in Ref.~\cite{DassalNG1} handled the case of evolution off a hard primary dipole in the leading $N_c$ limit. The result for the non-global contribution $S(t)$ was obtained by dividing the resummed result by the contribution of primary emissions alone, off the hard emitting dipole. For our work here we have a situation with several possible emitting dipoles. In this situation one has to resort to the large $N_c$ limit in which one can treat the problem as independent evolution of only the {\emph{leading colour-connected dipoles in the hard process}}. Detailed formulae can be found in Ref.~\cite{BMS}.

In the Z+jet case, which is simpler, we have noted that there is no visible difference arising from considering just the leading colour-connected dipoles in 
the hard process relative to the case where one evolves \emph{all} hard dipoles using the evolution code. We choose the latter option here hence write the full resummed expression, with the exception of the contributions coming from the constant terms $C_1$, as
\beq \label{masterZj}
\Sigma(v)= \sum_\delta \int d \mathcal{B} \frac{d \sigma^{(\delta)}_0}{d \mathcal{B}}  f^{(\delta)}_{\mathcal{B},{\rm global}} f^{(\delta)}_{\mathcal{B}, {\rm non-global}} \mathcal{H(B)}.
\eeq
The resummation of non-global logarithms, including contributions from the colour suppressed hard dipoles, is encoded in the two terms
\begin{eqnarray} \label{NGZjet}
f^{(\delta_1)}_{\rm non-global}&=& \exp \left (-C_A  C_F I_{13}(R) f_{13}(t) -\frac{C_A^2}{2} I_{12}(R )f_{12}(t)   \right), \\
f^{(\delta_2)}_{\rm non-global}&=& \exp \left (-C_A^2 I_{13} (R)f_{13}(t)-C_A \left( C_F -\frac{C_A}{2}\right) I_{12}( R ) f_{12}(t)  \right),
\end{eqnarray}
where we have used that $I_{13}=I_{23}$; the other contribution $I_{12}$ does depend on $R$ and vanishes in the $R\to 0$ limit, where one recovers the picture of jets evolving independently. As we stated before, taking the large $N_c$ limit of the non-global contributions would amount to switching off the contribution from the colour suppressed dipoles or, equivalently, choosing $C_F =C_A/2$ in the above. This produces no significant difference to our results but the result written above has the advantage that it includes the the full contribution to $\ord(\as^2)$ non-global coefficient.
We have defined
\begin{equation}
f_{ij}(t)=  \frac{1+(a_{ij} t)^2}{1+(b_{ij}t)^{c_{ij}} } t^2\,, \quad t=\frac{1}{4 \pi \beta_0} \ln \left(1-2 \as(p_T) \beta_0 \ln\frac{R^2}{v}\right)
\end{equation}
The coefficients  $a_{ij}, b_{ij}, c_{ij}$ are obtained by fitting the functional form above to the numerical results from the large-$N_c$ dipole-evolution code. Numerical results are reported in Appendix~\ref{app:resum}.

Following a similar method, we obtain the corresponding result for the dijet case:
\bea \label{masterdijets}
\Sigma(v)&=& \sum_{\delta,J=3,4} \int d \mathcal{B} \,{\rm tr} \left[ \frac{ H_{\delta} e^{-\left( \mathcal{G}_{\delta,J}+\gamma_E \mathcal{G}'_{\delta,J}+ \mathcal{S}_{\delta,J}\right)^{\dagger}} e^{-\mathcal{G}_{\delta,J}-\gamma_E \mathcal{G}'_{\delta,J}-\mathcal{S}_{\delta,J}}+(\Delta y \leftrightarrow - \Delta y)  }{\Gamma\left(1+ 2 \mathcal{G}'_{\delta,J} \right)}  \right]\mathcal{H(B)}.\nonumber \\
\eea
Up to small $\ord\left(R^4 \right)$ corrections, we have
\begin{eqnarray}
\mathcal{S}_{\delta,3}&=&  \Big [ \frac{C_A}{2} \Big (I_{13}(R){\bf T}_3^2 f_{13}(t)-  {\bf T}_1.{\bf T}_2 I_{12} ( R )f_{12}(t)-
 {\bf T}_1 . {\bf T}_4  I_{14}(R,\Delta y) f_{14}(t)   \nonumber \\ &&-{\bf T}_2.{\bf T}_4  I_{24}(R, \Delta y)f_{24}(t) \Big) \Big],\nonumber \\
\mathcal{S}_{\delta,4}&=& \Big [ \frac{C_A}{2} \Big (I_{13}(R){\bf T}_4^2 f_{13}(t)-  {\bf T}_1.{\bf T}_2 I_{12} ( R )f_{12}(t)-
 {\bf T}_2 . {\bf T}_3  I_{14}(R,\Delta y) f_{14}(t) \nonumber \\ &&  -{\bf T}_1. {\bf T}_3  I_{24}(R, \Delta y)f_{24}(t) \Big) \Big],
\end{eqnarray}
where we have used that $I_{13}=I_{23}\simeq I_{34}$.
As before the above expressions capture the full colour structure of the non-global contribution at $\ord(\as^2)$, but beyond that are valid only in the large-$N_c$ limit.

\section{Z+jet at the LHC} \label{sec:Zjet}
In this section we investigate the numerical impact of the different contributions which are relevant in order to achieve NLL accuracy. We decide to study the differential distribution
$\frac{d \sigma}{d \zeta} $, where $\zeta =\sqrt{v}=\frac{m_{ J}}{p_{ T}}$, so as to study the jet mass distribution directly rather than the squared jet mass. We also find it useful to work with a dimensionless ratio to better separate soft physics contributions. In fact, a fairly large value of the jet mass can be generated by the emission of a very soft gluon, if the transverse momentum of the hard jet is large, while small values of $\zeta$ always correspond to the emission of soft and/or collinear gluons. If not stated otherwise, we normalise our curves to the Born cross-section. We use the matrix element generator \textsc{Comix}~\cite{comix}, included in \textsc{Sherpa}~\cite{sherpa}, to produce all the tree-level cross-sections and distributions. We consider proton-proton collisions at $\sqrt{s}=7$~TeV and  we select events requiring $p_{T}>200$~GeV;  the Z boson is produced on-shell and does not decay. Jets are defined using the anti-$k_t$ algorithm~\cite{antikt}.
In our calculation we use the set of parton distribution function \textsc{Cteq}6m~\cite{cteq6m}, with renormalisation and factorisation scales fixed at $\mu_R=\mu_F=200$~GeV, to ease the comparison with different Monte Carlo parton showers, which we are going to perform in Section~\ref{sec:showers}. 

\subsection{Different approximations to the resummed exponent} \label{sec:exponent}

\begin{figure} 
\begin{center}
\includegraphics[width=0.7\textwidth]{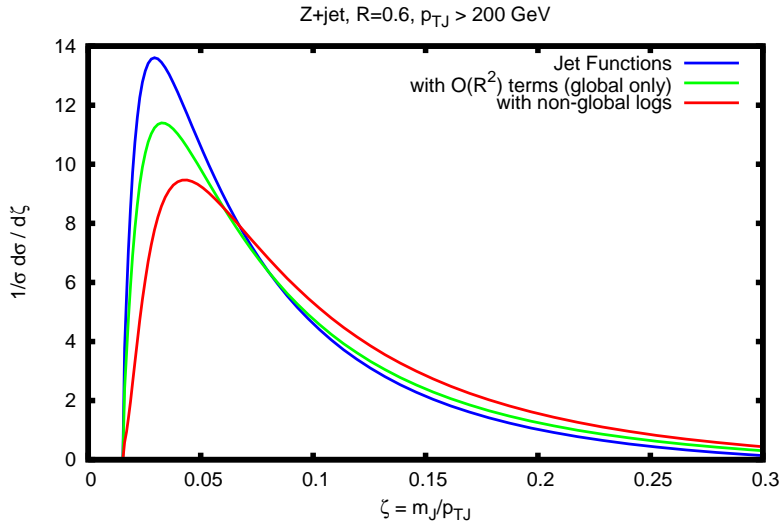}
\caption{Comparison between different approximations to the resummed exponent: jet functions (blue), with full resummation of the global contribution (green) and with non-global logarithms as well (red). The jet radius is $R=0.6$.} \label{fig:comparisonR06}
\end{center}
\end{figure}

\begin{figure} 
\begin{center}
\includegraphics[width=0.7\textwidth]{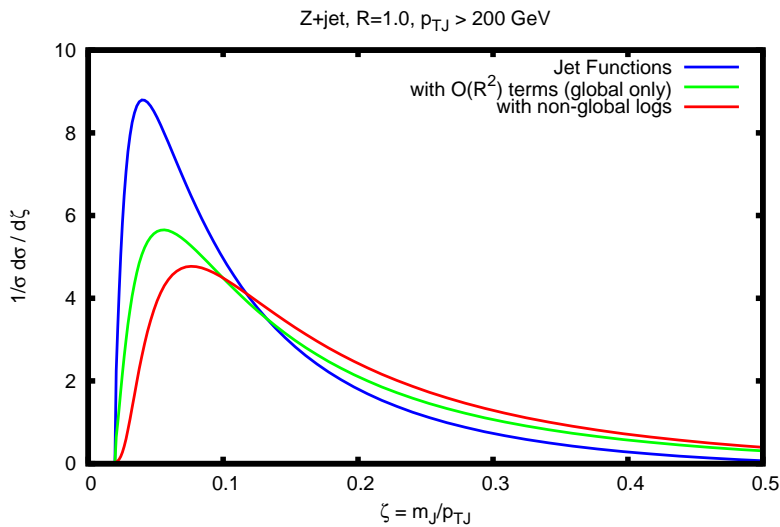}
\caption{Comparison between different approximations to the resummed exponent: jet functions (blue), with full resummation of the global contribution (green) and with non-global logarithms as well (red). The jet radius is $R=1.0$.} \label{fig:comparisonR10}
\end{center}
\end{figure}

We start by considering different approximations to the resummed exponent $f^{(\delta)}_{\mathcal{B}}$. 
We present our results for two different jet-radii: $R=0.6$, in Fig.~\ref{fig:comparisonR06},  and $R=1.0$, in Fig.~\ref{fig:comparisonR10}.
The blue curve corresponds to the most simple approximation to the NLL resummed exponent: the jet-function. This approximation correctly resums soft and collinear radiation as well as hard collinear, but does not capture all soft radiation at large angle. In particular this corresponds to neglecting terms that are suppressed by powers of the jet radius in the resummed exponent Eq.~(\ref{Zjet_rad}). These terms are included in the resummation of all global contributions (green line). We have checked that inclusion of $\ord(R^2)$ is enough because the $\ord(R^4)$ corrections are below the percent level even for $R=1.0$.  We stress that up to this point no approximation on the colour structure has been made, although we have checked that sub-leading colour corrections are small, once the collinear part has been properly treated. Finally, in the red curve we also take into account the resummation of non-global logarithms as described by Eq.~(\ref{masterZj}). The first, $\ord(\as^2)$, coefficient on the non-global contribution is computed exactly, while the subsequent resummation is performed 
using a numerical dipole-evolution code in the large-$N_c$ limit. 
We note that the inclusion of $\ord(R^2)$ terms in the resummed exponent as well as non-global logarithms, noticeably corrects the simple jet-function picture, based on collinear evolution. The peak height is reduced by more than 30\% for $R=0.6$ and it is nearly halved for $R=1.0$. The effect of non-global logarithms is reduced in the latter case, but the $\ord(R^2)$ corrections to the jet-function approximation become bigger.

 \FloatBarrier

\subsection{Matching to fixed order} \label{sec:matching}
\begin{figure} 
\begin{center}
\includegraphics[width=0.49\textwidth]{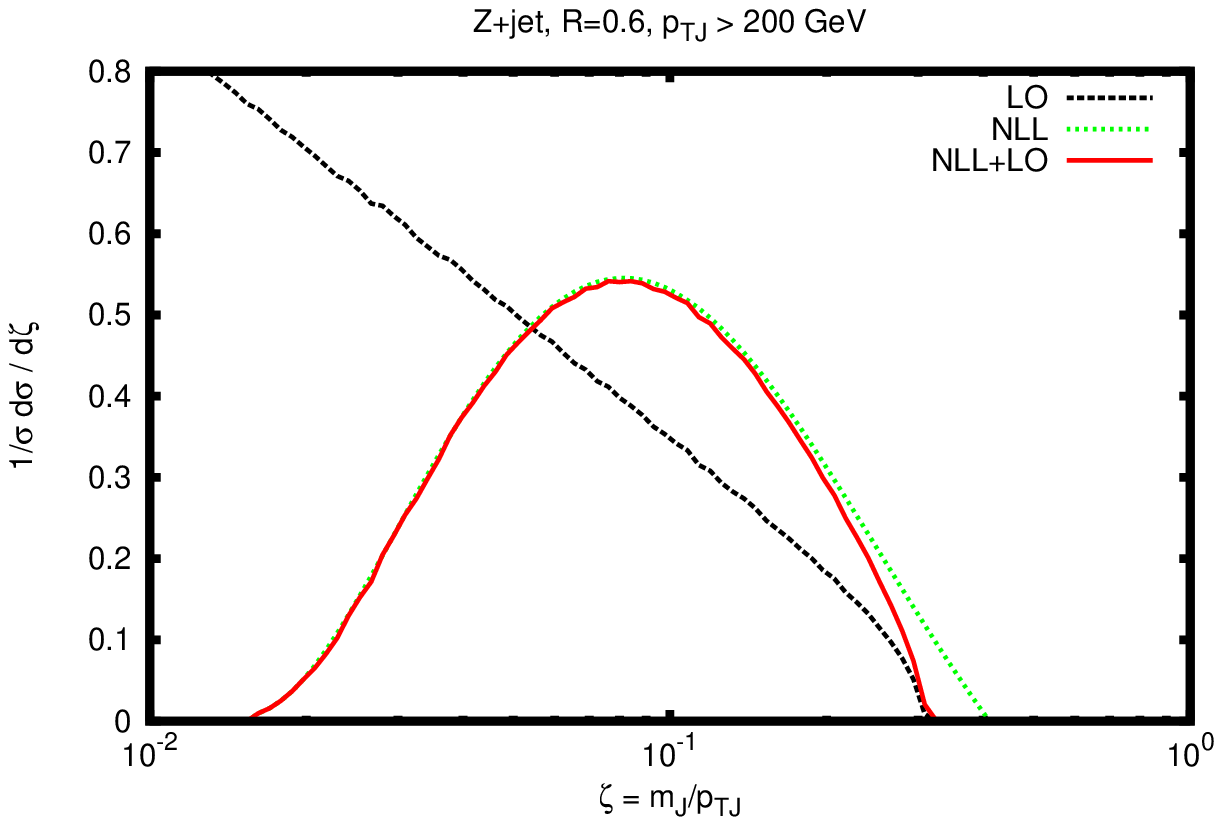}
\includegraphics[width=0.49\textwidth]{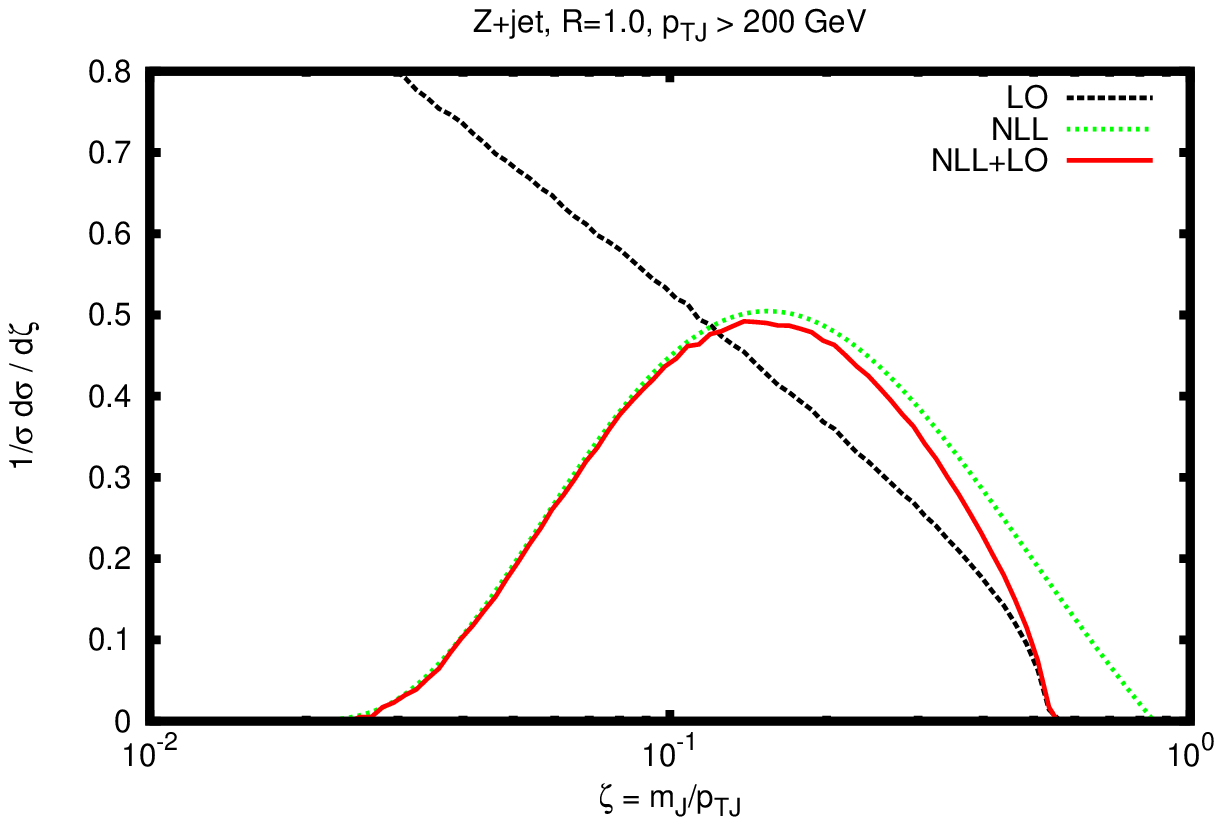} 
\caption{Matching of the NLL resummed distribution to the LO one for $R=0.6$ (on the left) and $R=1.0$ (on the right).}\label{fig:matching}
\end{center}
\end{figure}

We now turn our attention to obtaining a jet-mass distribution which is reliable for all values of $\zeta$. We achieve this by matching to a fixed-order (FO) calculation. 
Although a complete phenomenological analysis would require matching to a next-to leading order (NLO) QCD calculation, for the purpose of this paper we limit ourselves to LO matching. 
We compute the differential jet-mass distribution at $\ord(\as)$ using \textsc{Comix}. The result is plotted in Fig.~\ref{fig:matching} (dashed black line): the differential distribution $\frac{d \sigma}{d \ln \zeta}$ diverges logarithmically in the small-mass limit.  The dotted green line instead represents the resummed result. The matched curve (shown in solid red) is obtained by straightforwardly adding the two contributions and removing the double-counted terms, i.e. the expansion of the resummation to $\ord(\as)$:
\beq \label{matching}
\frac{1}{\sigma}\frac{{ d \sigma}_{\rm NLL+LO}}{d \ln \zeta}= \frac{1}{\sigma}\left[ \frac{d \sigma_{\rm LO}}{d \ln \zeta}+\frac{d \sigma_{\rm NLL}}{d \ln \zeta}-\frac{d \sigma_{{\rm NLL},\as}}{d \ln \zeta} \right].
\eeq
We note that the matched result coincides with the resummation at small $\zeta$, because the logarithmically divergent contributions to the LO distribution are cancelled by the expansion of the resummation. Moreover, the matched distribution follows the LO order one in the opposite limit. In particular, we note that the LO result exhibits an end-point:
\begin{equation}
\zeta^2 =\frac{m_{ J}^2}{p_{T}^2} =  \frac{2 p_t k_t}{|\underline{p}_t+\underline{k}_t|^2} \left(\cosh y - \cos \phi \right)=\frac{2 p_t k_t}{p_t^2+k_t^2+2 p_t k_t \cos \phi} \left(\cosh y - \cos \phi \right),
\end{equation}
which leads to 
\begin{equation}
\zeta_{\rm max}=\sqrt{\max_{y^2+\phi^2\le R^2} \zeta^2}=\tan \frac{R}{2}=\frac{R}{2}+O(R^3).
\end{equation}
This feature is not present in the resummed distribution, because the jet does not recoil against the emission of the eikonal gluon, but it is restored thanks to matching.

 \FloatBarrier

\subsection{Numerical estimate of constant term $C_1$} \label{sec:C1num}

In this section we investigate the numerical impact of the constant $C_1$ defined in Eq.~(\ref{C1def}), in the case of Z+jet events, where we measure the mass of the highest $p_{\rm T}$ jet.
Different contributions build up the constant terms at $\mathcal{O}(\alpha_s)$: we have one-loop virtual corrections, terms that cancel the renormalisation and  factorisation scale dependence of the Born pieces and contributions with multiple partons in final state that may, or may not, end up in the same jet. 
One-loop virtual corrections have the same kinematical and flavour structure as the corresponding Born subprocess $\delta$ and, consequently, they need to be suppressed by the same resummed exponent. In order to cancel infrared divergencies, one needs to consider real emissions in the soft and collinear limit as well. The final-state singularities are precisely the source of the logarithms we are resumming and these configurations can be mapped  onto one of the Born subprocesses $(\delta)$. Initial-state collinear singularities, which do not give rise to any logarithms of the jet mass, must be absorbed into the parton densities, leaving behind terms that depend on the factorisation scale. Finally, we also need to consider kinematic configurations where the two final-state partons are not recombined, resulting into Z+2 jet events, and suppress the hardest one with the appropriate exponent. Thus, we should perform the calculation of the NLO cross-section that appears in Eq.~(\ref{C1def}) keeping track of the kinematics of the final-state partons, in order to separate between $g$ and $q$ components. Although this can be clearly done by computing these corrections analytically, for the current analysis we use the program MCFM~\cite{mcfm}, which does not trivially allow us to do so. Nevertheless, we are able to compute the finite part of the virtual corrections, for the different initial-state channels. 
We suppress virtual corrections, as well as the integrated term in Eq.~(\ref{C1def})  with their appropriate form factor, $f^{(\delta)}$. We then multiply all the remaining real corrections by either a gluon or a quark form factor, producing a band. 

\begin{figure} 
\begin{center}
\includegraphics[width=0.49\textwidth]{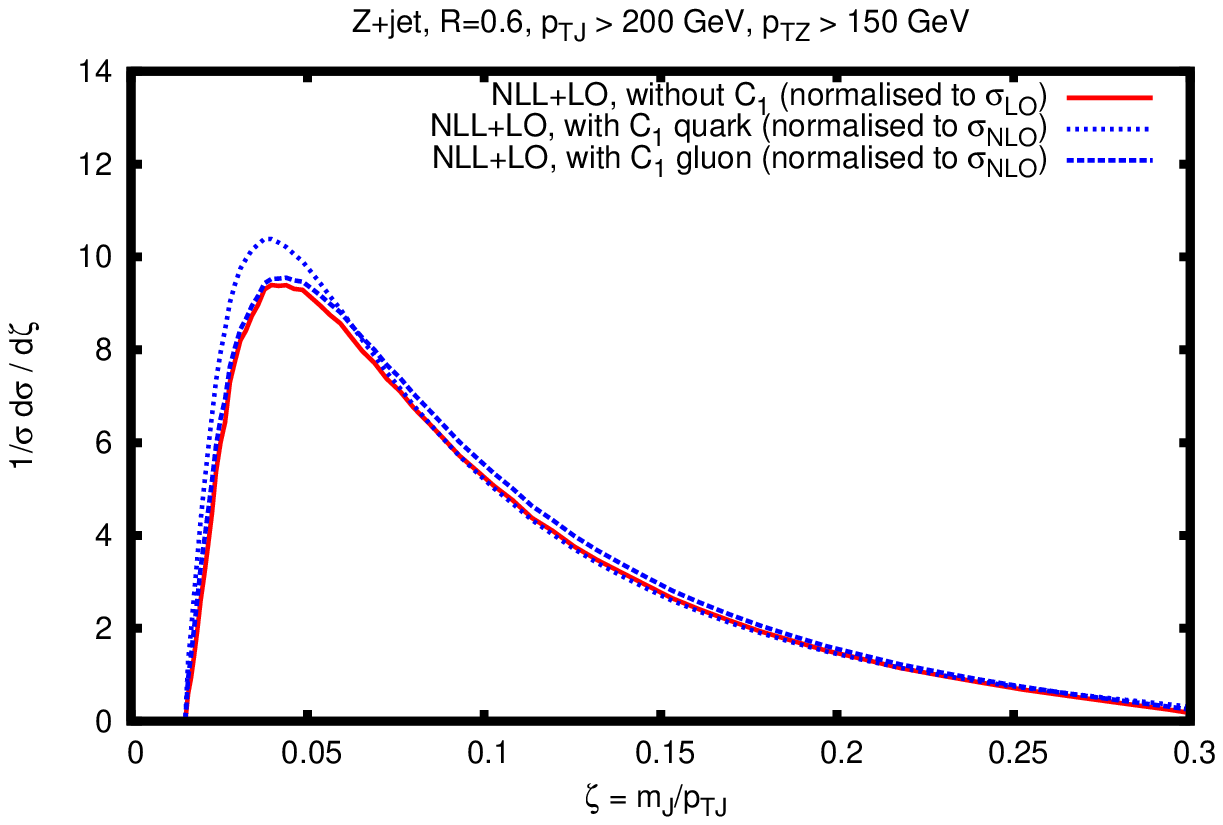}
\includegraphics[width=0.49\textwidth]{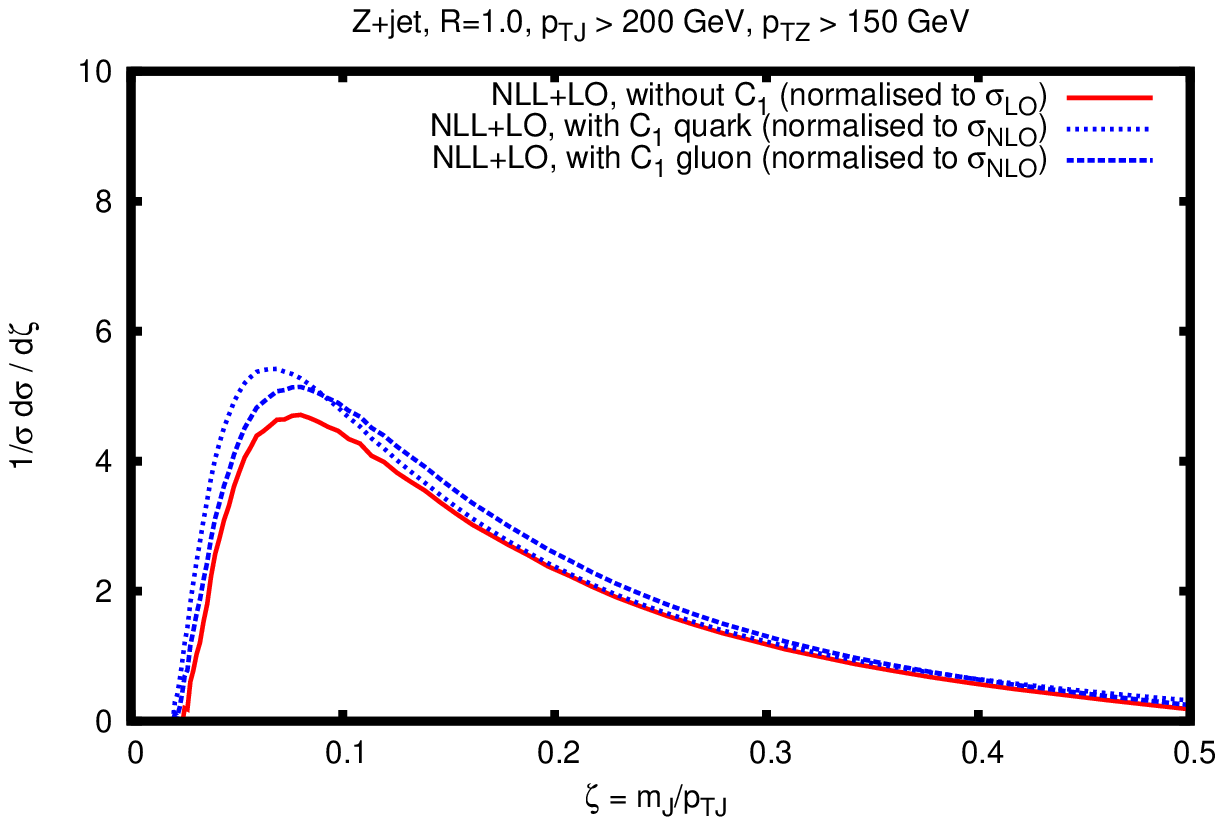}
\caption{The impact of the NLL constant $C_1$, for $R=0.6$ jets (on the left) and $R=1.0$ (on the right). The band is produced by suppressing the real radiation contributions with a quark or gluon jet form factor, as explained in the text.}\label{fig:C1}
\end{center}
\end{figure}

Our findings are plotted in Fig.~\ref{fig:C1}, for $R=0.6$, on the left, and $R=1.0$, on the right\footnote{We have suppressed $C_1$ with the full global resummed exponent, producing terms beyond our NLL accuracy, which we do not control. A more precise analysis would involve the complete determination of $C_1$, suppressed only with double-logarithmic terms, together with an uncertainty band, assessing the impact of higher logarithmic orders.}.
When $C_1$ is included, we normalise the distribution to the total NLO rate, rather than the usual Born cross-section. In order to avoid large NLO corrections~\cite{giantKfact}, we put a cut on the Z boson transverse momentum $p_{ TZ}>150$~GeV.  We have found that this leads to $K$-factor $K=1.45$ and $K=1.57$, for $R=0.6$ and $R=1.0$, respectively.  With this set of cuts, the impact of $C_1$ is not too big, but definitely not negligible. The complete calculation of this constant is therefore necessary in order to be able to perform accurate phenomenology. 

 \FloatBarrier
\subsection{Comparison to Monte Carlo event generators}\label{sec:showers}
In this section we compare our resummed and matched result NLL+LO to three standard Monte Carlo event generators: \textsc{Sherpa}~\cite{sherpa},  \textsc{Pythia}~\cite{pythia} and \textsc{Herwig}{\footnotesize ++}~\cite{herwig}. 
Monte Carlo parton showers are powerful tools to simulate complicated final states in particle collisions. When interfaced with hadronisation models, they are able to describe the transition between partons and hadrons. Moreover, they 
provide events which are fully differential in the particles' momenta. Also 
they provide models of other non-perturbative effects at hadron colliders such as the underlying event where it is not possible to have phenomenological estimates from first principles of QCD. As stated before however, despite their usefulness and successes, it is quite difficult to assess the theoretical precision of these tools. For this reason, comparisons between parton showers and analytic calculations, which have a well-defined theoretical accuracy, form an important part of QCD phenomenology. For the case of jet-mass it has been noted in recent ATLAS studies~\cite{ATLASjetmass} for jet masses that \textsc{Pythia} with hadronisation and the underlying event gives a reasonably good description of the data. We would therefore expect our resummation to be in some accordance with \textsc{Pythia} though we should stress that we do not include any non-perturbative effects. Hence we compare our results to the parton shower aspect of the various event generators on its own. While this is in principle correct, in practice one should be aware that there can be considerable interplay between the shower level and the non-perturbative effects in various event generators so that these programs may only return more meaningful results when all physical effects (perturbative and non-perturbative) are considered together. We should bear this caveat in mind while attempting to compare a resummed prediction with just the parton shower models in event generators. 

\begin{figure} 
\begin{center}
\includegraphics[width=0.7\textwidth]{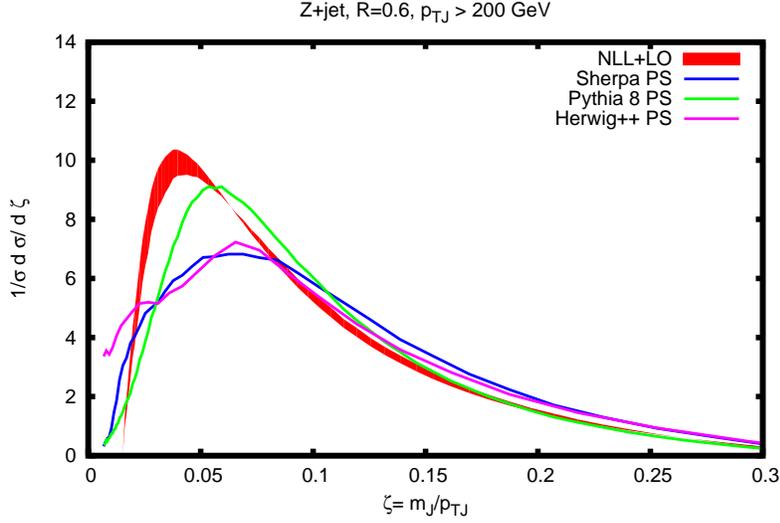}
\caption{Comparison of our resummed and matched result NLL+LO  (in red) to standard Monte Carlo event generators, at the parton level.} \label{fig:shower}
\end{center}
\end{figure}

The results of the comparison are shown in Fig.~\ref{fig:shower}.  Our NLL+LO result for $R=0.6$ is shown in red (the band represents the uncertainty due to the incomplete determination of $C_1$). The Monte Carlo results are obtained  with the same parton densities as in the resummed calculation and the same set of cuts. For \textsc{Sherpa} (blue line) and  \textsc{Pythia} (green line) we fix $\mu_R=\mu_F=200$~GeV, while for \textsc{Herwig}{\footnotesize ++} (magenta line) we use the default transverse mass of the Z boson. 
At the shower level, \textsc{Sherpa} and  \textsc{Herwig}{\footnotesize ++} appear to perform quite similarly. They produce fairly broad distributions, which 
are not very much suppressed as $\zeta \to 0$. 
The distribution obtained from \textsc{Pythia} instead produces a curve which is much closer to our resummed result. Although the position of the peaks differ by $ \delta \zeta =0.01$ ($\delta m_{\rm J}\sim2$~GeV), the height and general shapes appear in agreement. 

\begin{figure} 
\begin{center}
\includegraphics[width=0.7\textwidth]{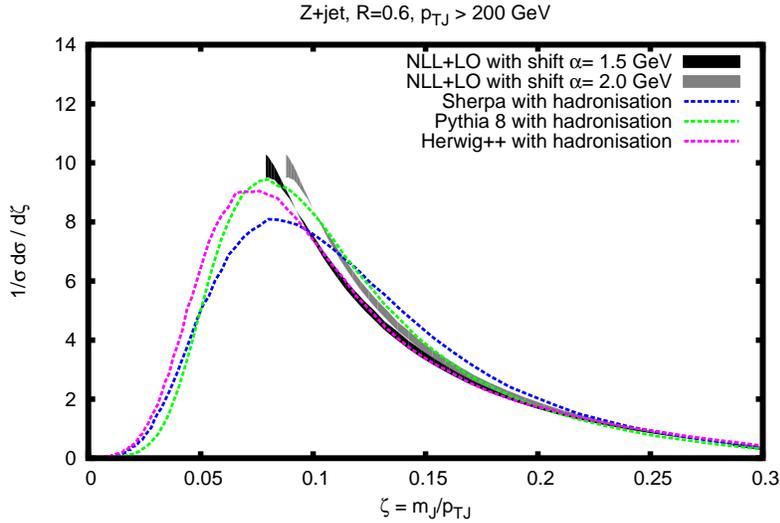}
\caption{Results for the $\zeta$ distributions obtained with standard Monte Carlo parton showers, with hadronisation corrections (dashed lines) compared to analytical resummation with non-perturbative shifts (shaded bands) as explained in the main text.} \label{fig:hadro}
\end{center}
\end{figure}

The agreement between the different Monte Carlo generators is restored when hadronisation corrections are included, as demonstrated in Fig.~\ref{fig:hadro}:  \textsc{Pythia} and \textsc{Herwig}{\footnotesize ++} produce very similar results, while the distribution obtained with \textsc{Sherpa} is broader, but not too different. Clearly, in order to compare to collider data, one must also include the contribution from the underlying event. 

It is also interesting to compare the resummed prediction we computed in this paper with a shift to the right to account for hadronisation corrections to the event generator results after hadronisation. The shift approximation, initially suggested in \cite{DokWeb97}, should be valid to the right of the Sudakov peak but will certainly break down in the vicinity of the peak (see also the discussion in \cite{DasSalDIS} and references therein). The amount of the shift is related to the non-perturbative correction to mean values of jet masses derived in \cite{Dassalmag}. From that reference, we note that the $v=m_J^2/p_T^2$ distribution should be shifted by an amount $\alpha R/p_T$ (the correction to the mean value) with a dimensionful coefficient $\alpha$ one can take to be of order of a few times $\Lambda_{\rm QCD}$. The results of our calculations with a non-perturbative shift are shown in Fig.~\ref{fig:hadro}. From there one notes that a shift of $1.5\,\mathrm{GeV}\, R/p_T$ (for the $v$ distribution which we carry over to the $\zeta$ distribution plotted in Fig.~\ref{fig:hadro}) where we take $p_T$ to be the value of the lower bound (200 GeV in this case) on transverse momentum, yields an excellent agreement with the \textsc{Herwig}++ result after hadronisation. On the other hand, a slightly larger shift of $2.0 \, \mathrm{GeV}\, R/p_T $ yields a good agreement with \textsc{Pythia}. We have truncated the shifted resummed result near the peak of the resummed distribution, as we would expect the shift to not be meaningful beyond this region at the very best. Although we have made a crude estimate of hadronisation effects, and one may be able to better compute these corrections, we do note that within non-perturbative uncertainties our results are compatible with the most widely used event generator models. We therefore anticipate that after the further improvements we have in mind are accomplished, our results may be directly used for phenomenological purposes. 

 \FloatBarrier
\section{Dijets at the LHC} \label{sec:dijets}

\begin{figure} 
\begin{center}
\includegraphics[width=0.7\textwidth]{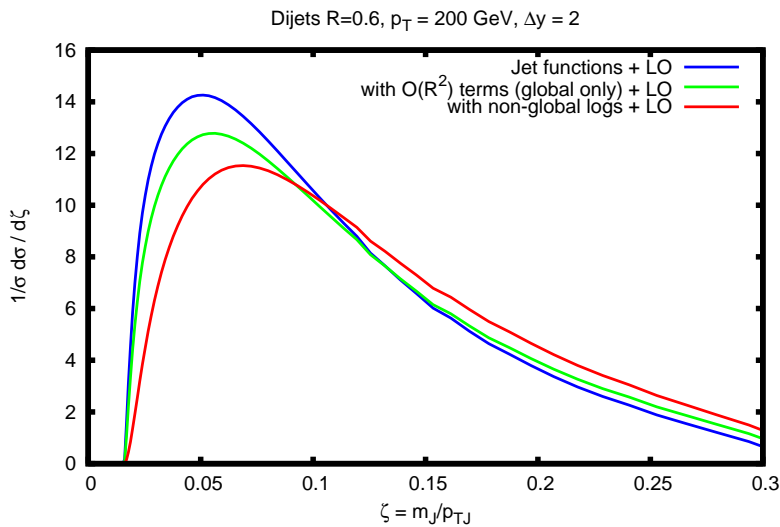}
\caption{The NLL+LO jet mass distribution for dijets, with different approximation to the resummed exponent.}\label{fig:dijets}
\end{center}
\end{figure}

In this section we provide numerical predictions for the jet mass distribution in dijet events. As before, we consider proton-proton collision at $\sqrt{s}=7$~TeV, with jets defined according to the anti-$k_t$ algorithm~\cite{antikt}.
The main complication with respect to the Z+jet case previously discussed is a more complicated colour algebra, which results into a matrix structure of large-angle soft gluon radiation. In order to simplify our resummed calculation, we work at fixed kinematics, i.e. we demand the jets' transverse momentum to be $p_{T}=200$~GeV and their rapidity separation to be $|\Delta y| = 2$ (at Born level we only have two jets).  We remind the reader that we consider
\beq
\frac{1}{\sigma}\frac{{ d \sigma}}{d  \zeta}= \frac{1}{\sigma} \left( \frac{d \sigma}{d \zeta_1} +\frac{d \sigma}{d \zeta_2}\right)_{\zeta_1=\zeta_2=\zeta}.
\eeq
We match the resummation to a LO calculation of the jet mass distribution obtained with \textsc{Nlojet}{\footnotesize++}~\cite{nlojet}, according to Eq~(\ref{matching}). In Fig.~\ref{fig:dijets} we plot our NLL+LO result with different approximation for the resummed exponent Eq.~(\ref{masterdijets}): jet-functions (blue), with the inclusion of $\ord(R^2)$ corrections (green) and non-global logarithms (red), as explained in detail for the Z+jet case. Although the corrections to the jet-function approximation are less pronounced than in the Z+jet case, they are still sizeable and must be taken into account. In our understanding, the perturbative part of the resummed result of Ref~\cite{yuan1,yuan2} is precisely the one captured by the jet-functions (plus an approximated treatment of the one-loop constant $C_1$).  

In order to obtain a more realistic prediction, one would need to integrate over the appropriate cuts in transverse momentum and rapidity and match to NLO, which, in principle, should not pose any difficulties. More delicate is instead the determination of the constant $C_1$, although this issue has been addressed in Ref.~\cite{BSZcaesar, BSZpheno} for the case of global event-shapes.

\section{Conclusions and Outlook}\label{sec:conclusions}

In this paper we have provided resummed expressions to NLL accuracy for jet-mass distributions in hadron-hadron collisions, for jets of arbitrary radius and defined in the anti-$k_t$ algorithm. In particular, we have considered Z boson production in association with a jet and jet masses in dijet production. We have improved upon existing studies of jet masses and shapes (many of which use $e^{+}e^{-}$ annihilation as a model for jets produced in hadron collision processes) by incorporating initial state radiation effects as well as accounting for non-global logarithms in the large $N_c$ limit. We have matched our results to leading-order predictions to account for non-logarithmic terms to this order and have commented on the role of the coefficient function $C_1 \alpha_s$ (corrections of order $\alpha_s$ to the basic resummation off the Born configuration). Finally, we have compared our results to all the leading Monte Carlo event generators  \textsc{Pythia}, \textsc{Sherpa} and \textsc{Herwig}{\footnotesize++}, with and without non-perturbative hadronisation corrections. 

We have found that firstly ISR and non-global logarithms play an important role even at relatively small values of jet radius such as $R=0.6$ and hence cannot be ignored while discussing inclusive jet shapes at hadron colliders. Our calculations for non-global logarithms both at fixed-order and at all orders represent to our knowledge the first attempt to calculate these terms beyond the simpler cases of $e^{+}e^{-}$ annihilation and DIS. On comparing our results to event generators widely used for phenomenology we find that at the purely perturbative level the best agreement is with the \textsc{Pythia} shower, with an apparent small shift accounting for much of the difference between the analytical resummation and the parton shower estimate. The differences with  \textsc{Sherpa} and \textsc{Herwig}{\footnotesize++} on the other hand are more marked especially towards smaller jet masses as one approaches the peak of the distributions. After hadronisation corrections are applied in the event generators and also applying a shift to the analytical resummation to account for hadronisation, we are able to obtain very good agreement with  \textsc{Pythia} and \textsc{Herwig}{\footnotesize++} with slightly different shifts required in either case, for jet mass values to the right of the Sudakov peak of the distribution. For smaller jet masses we do not expect a simple shift of the analytical resummation to reproduce the correct result and we observe here discrepancy with all Monte Carlo generators which is to be expected. However, this region of very small jet masses of the order of a few GeV will not be of interest in LHC phenomenology in any case. 
We may thus expect that our results should be of direct phenomenological value even pending some of the improvements we intend to make in the near future. 

For the immediate future we aim to improve our results chiefly by taking proper account of the order $\alpha_s$ coefficient function $C_1$ and computing the various pieces of $C_1$ which originate from different regions and suppressing these by an appropriate form factor rather than the cruder treatment reported in the text, which was aimed at producing an uncertainty band associated to a lack of correct treatment of $C_1$. When this is done we will be in a position to carry out 
an NLO matching and estimate the uncertainty of our theoretical calculations accurately, which will be important in the context of phenomenology. 

We have not taken directly into account non-perturbative effects in this paper, incorporating them as a simple shift of the perturbative spectra as for the case of global event shapes~\cite{Dassalrev}. We can study non-perturbative corrections in more detail using the methods outlined in~\cite{Dassalmag}, but for the moment we note that the Monte Carlo event generators we have studied, contain differences in their estimate of hadronisation which should be explored further along with studying the underlying event (UE) contributions. Since our predictions are valid for any value of jet radius $R$, one can hope that phenomenological studies selecting a small value of $R$ would help in better isolating the perturbative contributions, since the hadronisation and UE corrections for $m_J^2/p_T^2$ scale respectively as $R$ and $R^4$~\cite{Dassalmag}. On the other hand moving to a larger $R$ after pinning down the perturbative content would help to more accurately constrain the non-perturbative models in various approaches.

Additionally, although in this article we have treated a single variable, the jet mass, it is actually straightforward to extend our treatment to the entire range of angularities as for instance were explored for jets in \cite{EHLVW1,EHLVW2}. The basic calculations we carried out here can be easily extended to include those variables as well as any variants of the jet mass itself such as studying the jet mass with an additional jet veto. Once we have accomplished an NLO matching we therefore intend to generalise our approach to accommodate a range of substructure variables in different hard processes. We hope that our work will eventually lead to improved estimates of the accuracy and hence more confidence in a detailed understanding of jet substructure  than is the case presently, which could in turn be important for a variety of substructure applications at the LHC.

\section*{Acknowledgements}
We wish to thank Andrea Banfi and Frank Krauss  for many useful discussions about resummation and parton showers.
In particular, we thank Peter Richardson and Marek Schoenherr for \textsc{Herwig}{\footnotesize++} and \textsc{Sherpa} support.
 MD would like to thank the IPPP for support via the IPPP associateship award which facilitated his visit to Durham and those of SM to Manchester, during the course of this work. 
This work is supported by UK's STFC. The work of MD is supported by the Lancaster-Manchester-Sheffield Consortium for Fundamental Physics, under STFC grant ST/J000418/1.

\appendix

\section{Dipole calculations for the global part} \label{app:global}
In order to carry out the calculations for the various dipole terms required for the resummation detailed in the main text we shall make use of the following 
Born kinematics for the hard incoming and outgoing particles,
\begin{eqnarray}
p_1 &=& \frac{\sqrt{s}}{2} x_1 \left (1,0,0,1 \right), \nonumber \\
p_2 &=& \frac{\sqrt{s}}{2} x_2 \left (1,0,0,-1 \right), \nonumber \\
p_3 &=& p_t \left( \cosh y, 1,0,\sinh y \right), \nonumber \\
p_4 &=& p_t \left(\cosh y_r,-1,0,\sinh y_r \right),
\end{eqnarray}
where, as in the main, text $p_1$ and $p_2$ denote the four-momenta of the incoming hard partons, $p_3$ that of the jet whose mass is being considered and $p_4$ the four-momentum of a recoiling jet. In the case of a final state vector boson instead of a recoiling jet the contribution from $p_4$ to the dipole calculations is of course absent as the Z boson is colour neutral. Further we have indicated by $x_1$ and $x_2$ the fractional momenta of incoming partons relative to the colliding hadrons $P_1$ and $P_2$ respectively. Also $y$ denotes the rapidity of the measured jet, $y_r$ that of the recoiling jet and $p_t$ their transverse momenta. One can, by ignoring recoil, use the Born kinematics 
outlined above even in the presence of a soft emission, whose momentum can be expressed as \beq k=  k_t \left(\cosh \eta, \cos \phi, \sin \phi, \sinh \eta \right); \eeq 
where transverse momentum $k_t$, rapidity $\eta$ and azimuth $\phi$ are all measured with respect to the beam direction. 

We shall now set up the various dipole contributions to the jet mass cross-section. We recall from the main text that eventually these dipole contributions will exponentiate with appropriate colour factors, so what follows below is in essence a calculation of various components of the resummed exponent.

\subsection{In-in dipole}
The in-in dipole is made from incoming partons $1$ and $2$ and provides the 
contribution 
\bea
\mathcal{R}_{12} &=& \int k_t dk_t d\eta \frac{d\phi}{2\pi} \frac{\alpha_s \left (\kappa_{t,12} \right) }{2 \pi} \frac{(p_1.p_2)}{(p_1.k)(p_2.k)}\Theta \left 
( \frac{2 k_t}{p_t} \left (\cosh (\eta-y) -\cos \phi \right) -v \right) \nonumber \\ &&
\Theta \left (R^2 -\left( (\eta-y)^2+\phi^2 \right) \right)
\eea
We next note that $\kappa^2_{t,12} =  2 \frac{(p_1.k)(p_2.k)}{(p_1.p_2 )} = k_t^2$ and carry out the integration over $\eta$ and $\phi$. In doing so we can neglect the dependence of the jet mass on $\eta,\phi$ and just retain its dependence on $k_t$ since the coefficient of $k_t$ which contains the $\eta,\phi$ dependence, produces terms that are below single logarithmic accuracy. Performing the integral over $\eta$ and $\phi$ over the interior of the jet region gives us to the relevant single-log accuracy that 
\begin{equation}
\mathcal{R}_{12} = {R^2} \int_{v p_t}^{p_t} \frac{\alpha_s (k_t)}{2 \pi} \frac{dk_t}{k_t},
\end{equation}
where the lower limit of integration stems from the constraint on the jet-mass variable. 
We note that the dipole consisting of the two incoming partons gives rise to a pure single-logarithmic behaviour as is evident from the fact that the emitted gluon is inside the jet region, away from the hard legs constituting the dipole and hence there are no collinear enhancements. The soft wide-angle single logarithm we obtain is accompanied by an $R^2$ dependence on jet radius, reflecting the integration over the jet interior.

\subsection{In-recoil dipoles}

As before for the in-in case, for a dipole made up of an incoming parton and one recoiling against a measured jet, we have that the emitted gluon is in the interior of the jet away from the emitting legs,  and produces no collinear logarithm. We have that 
\bea
\mathcal{R}_{14} &=& \int k_t dk_t d\eta \frac{d\phi}{2\pi} \frac{\alpha_s \left (\kappa_{t,14} \right) }{2 \pi} \frac{(p_1.p_4)}{(p_1.k)(p_4.k)}\Theta \left 
( \frac{2 k_t}{p_t} \left (\cosh (\eta-y) -\cos \phi \right) -v \right)  \nonumber \\ &&
\Theta \left (R^2 -\left( (\eta-y)^2+\phi^2 \right) \right).
\eea
However , in this case we have that 
\begin{equation}
\kappa_{t,14}^2 = \frac{2(p_1.k)(p_4.k)}{(p_1.p_4)} =2 k_t^2 e^{y_r-\eta} \left( \cosh(\eta-y_r)+\cos \phi \right).
\end{equation}
To single logarithmic accuracy, we can write the result into the form
\begin{equation}
\int_{v}^{1} \frac{dx}{x}\alpha_s (x p_t)  \int \frac{e^{\eta-y_r}}{\cosh(\eta-y_r)+\cos \phi} \Theta \left (R^2 -\left( (\eta-y)^2+\phi^2 \right) \right). 
\end{equation}
It is possible to evaluate the integral over $\eta,\phi$ as a power series in $R$, such that the overall result for this dipole reads
\begin{equation}
\int_v^1\frac{dx}{x}\frac{\alpha_s (x p_t)}{2 \pi}  \left( \frac{R^2}{2} \frac{e^{\Delta  y}}{\left(1+\cosh \Delta y \right)}+\frac{R^4}{32} {\mathrm{sech}}^4 \frac{\Delta y}{2}+\mathcal{O} \left(R^6 \right) \right)
\end{equation}
where $\Delta y = y- y_r$
For the $\mathcal{R}_{24}$ dipole one obtains the same result as above, with the replacement of $\Delta y$ by $-\Delta y$.

\subsection{Jet-recoil dipole}
For this dipole we have also an involvement of the parton that initiates the 
measured jet as one of the hard emitting legs. The consequent collinear singularity, in addition to the pole due to soft emission, gives rise to 
double-logarithmic contributions. In this case one needs to evaluate the argument of the running coupling more carefully whereas for soft wide-angle pieces, dealt with previously,  all that is important at our accuracy is that the argument of the running coupling is proportional to the energy of the soft gluon being considered. 

In the present case we start with the integral 
\bea
\label{eq:jetreco}
\mathcal{R}_{34} &=& \int k_t dk_t d\eta \frac{d\phi}{2\pi} \frac{\alpha_s \left (\kappa_{t,34} \right) }{2 \pi} \frac{(p_3.p_4)}{(p_3.k)(p_4.k)} \Theta \left 
( \frac{2 k_t}{p_t} \left (\cosh (\eta-y) -\cos \phi \right) -v \right) \nonumber \\ &&
\Theta \left (R^2 -\left( (\eta-y)^2+\phi^2 \right) \right)
\eea
where
\begin{equation}
\kappa_{t,34}^2 = \frac{2(p_3.k)(p_4.k)}{(p_3.p_4)} = 2 k_t^2 
\frac{\left(\cosh \left(y_r-\eta \right) +\cos \phi \right) \left(\cosh \left(y-\eta \right)+\cos \phi) \right)}{1+\cosh \Delta y}.
\end{equation}
Next we introduce the variables $r$ and $\theta$ such that $y-\eta = r \cos \theta$ and $\phi = r \sin \theta$ such that the in jet condition $(y-\eta)^2+\phi^2<R^2$, simply produces $r < R$, while $\theta$ takes values between $0$ and $2\pi$. 
Thinking of Eq.~\ref{eq:jetreco} as a power series in $r$ we can isolate the collinear-enhanced terms which produces double logarithms and separate it from 
terms that are finite as $r \to 0$, which upon integration produce terms that 
behave as powers of $R$ corresponding to soft wide-angle contributions.
The contribution $\mathcal{R}_{34}$ can thus be broken into terms that contain collinear enhancements (and hence leading double logarithms) and pure soft single-logarithmic pieces:
\begin{equation}
\mathcal{R}_{34} = \mathcal{R}_{34}^\mathrm{coll.}+\mathcal{R}_{34}^\mathrm{soft},
\end{equation}
where the pure soft terms can be expressed as a power series in $R$:
\begin{equation}
\mathcal{R}_{34}^\mathrm{soft} =  \int_v^1\frac{\alpha_s (x p_t)}{2 \pi} \frac{dx}{x}
\left( \frac{R^2}{4} \tanh^2 \frac{\Delta y}{2}+\frac{R^4}{1152} \mathrm{sech}^4 \frac{\Delta y}{2} \left(-5+\cosh \Delta y \right)^2+\mathcal{O}(R^6) \right),
\end{equation}
where we have neglected terms that vanish as $v \to 0$.

The collinear contributions require us to be more precise in treating the 
argument of the running coupling. In particular, in the limit of small $r$, which is the collinear region, we obtain
\begin{equation}
\label{eq:fix}
 \mathcal{R}_{34}^{\mathrm{coll.}} = \int_v^{R^2}\frac{dr^2}{r^2} \frac{d\theta}{2 \pi} 
\int_{\frac{v}{R^2}}^1 \frac{dx}{x} \frac{\alpha_s}{2 \pi} \left(x r p_t \right)
\end{equation}
where we have written the result in terms of a dimensionless energy fraction $x$ such that in the collinear limit $r \ll 1$, one has that $\kappa_{t,34}/(r p_t) = x$. The lower limits of the energy and angular integrals follow from the restriction on the jet mass to be greater than $v$, while the upper limit of the angular integral over $r^2$ corresponds to the jet boundary. 

It is straightforward to obtain the result corresponding to the fixed-coupling limit from Eq.~(\ref{eq:fix}), which gives: 
\begin{equation}
 \mathcal{R}_{34}^{\mathrm{coll.}}= \frac{\alpha_s}{2 \pi} \frac{1}{2} \ln^2 \frac{R^2}{v}.
\end{equation}

The running coupling result can be compared directly to that obtained in the $e^{+}e^{-}$ case in our previous work~\cite{BDKM}. In order to do this we introduce the transverse momentum with respect to the emitting jet $k_{t,J} = x r p_t$, and changing variable we obtain

\begin{equation}
 \mathcal{R}_{34}^{\mathrm{coll.}}= \int \frac{dk_{t,J}}{k_{t,J}} \frac{\alpha_s \left( k_{t,J} \right)}{2 \pi} 
 \frac{dr^2}{r^2} \Theta \left ( R^2-r^2\right) \Theta \left (r^2-v \right) \Theta \left (r^2 -\frac{k^2_{t,J}}{p_t^2} \right) \Theta \left (r^2 -\frac{v^2 p_t^2}{k_{t,J}^2} \right)
\end{equation}

Carrying out the integral over $r$, we are left to evaluate the integral 
over $k_{t,J}$ which reads 

\begin{multline}
 \mathcal{R}_{34}^{\mathrm{coll.}} = \int \frac{d k_{t,J}^2}{k_{t,J}^2}  \frac{\alpha_s \left( k_{t,J} \right)}{2\pi}  \ln \left (\frac{R p_t}{k_{t,J}}   \right) \Theta \left (\frac{k_{t,J}^2}{p_t^2} -v \right) \Theta \left(R^2 -\frac{k_{t,J}^2}{p_t^2} \right)  \\ + \int  \frac{d k_{t,J}^2}{k_{t,J}^2} \frac{\alpha_s \left( k_{t,J}^2\right)}{2\pi} \ln \left (\frac{R k_{t,J}}{v p_{t}}   \right) \Theta \left (v-\frac{k_{t,J}^2}{p_t^2} \right ) \Theta \left(\frac{k_{t,J}^2}{p_t^2} -\frac{v^2}{R^2}\right)
\end{multline}

\subsection{In-jet dipoles}
Now we consider the dipole formed by the triggered jet and one of the incoming partons. Since here too the measured jet is involved as one of the legs of the hard dipoles, we can carry out precisely the same manipulations as for the $\mathcal{R}_{34}$ 
dipole immediately above. Doing so yields the same soft-collinear piece as before 
\begin{equation}
 \mathcal{R}_{13}^{\mathrm{coll.}}= \mathcal{R}_{23}^{\mathrm{coll.}}=\mathcal{R}_{34}^{\mathrm{coll.}},
\end{equation}
while the soft wide-angle piece reads
\begin{equation}
\mathcal{R}_{13}^{\mathrm{soft}} = \mathcal{R}_{23}^{\mathrm{soft}} = \int_{v}^{1} \frac{\alpha_s(x p_t)}{2 \pi} \frac{dx}{x} \, \left( \frac{R^2}{4}+\frac{R^4}{288}+\mathcal{O}\left(R^6 \right) \right).
\end{equation}

\section{Dipole calculations for the non-global contribution} \label{app:nonglobal}
We now turn our attention to the evaluation of the different contributions arising from correlated-gluon emissions.

\subsection{Dipoles involving the measured jet}
We start by considering the the three dipoles involving the triggered jet $I_{13}=I_{23}$: 
\bea
I_{13}&=& 4\int d\eta_1 \frac{d\phi_1}{2\pi} \int d \eta_2 \frac{d\phi_2}{2\pi} \nonumber \\ && \cdot
\frac{1-e^{\eta_1}\cos\phi_1-e^{\eta_2}\cos\phi_2+e^{\eta_1+\eta_2}\cos(\phi_1-\phi_2)}{\left(\cosh(\eta_1-\eta_2)-\cos(\phi_1-\phi_2)\right)\left(\cosh\eta_1-\cos\phi_1\right)\left(\cosh\eta_2-\cos\phi_2\right)} \nonumber \\ && \cdot \Theta \left(R^2-(\eta_2^2+\phi_2^2)\right) \Theta \left(\eta_1^2+\phi_1^2-R^2 \right).
\eea
We evaluate the above integral in powers of the jet radius $R$. We find
\beq \label{NG_I13num}
I_{13}=I_{23}= \frac{\pi^2}{3}+ \alpha_2 R^2+\alpha_4 R^4 +\ord \left(R^6  \right).
\eeq
As expected the $\ord\left(R^0 \right)$ term is the same result as in the hemisphere case. Moreover, we find that the  $\ord\left(R^2 \right)$ is compatible with zero and the first non-vanishing correction to $\frac{\pi^2}{3}$ appears to be  $\ord\left(R^4 \right)$. A numerical evaluation of the integral leads to small coefficient $\alpha_4\approx 0.013$.

The dipole $I_{34}$, which is relevant for the dijet calculation, depends, in principle, on the rapidity separation $\Delta y$ between the leading jets. 
This dependence will be associated with powers of $R$ that appear to make a negligible contribution. 
Its numerical determination gives a result consistent with Eq.~(\ref{NG_I13num}).


\subsection{The remaining dipoles}
Let us consider the dipole from 2 incoming legs $I_{12}$.

We have to evaluate 
\bea
I_{12}&=& 4\int d\eta_1 \frac{d\phi_1}{2\pi} \int d \eta_2 \frac{d\phi_2}{2\pi} 
\frac{\cos(\phi_1-\phi_2)}{\cosh(\eta_1-\eta_2)-\cos(\phi_1-\phi_2)} \Theta \left(R^2-(\eta_2^2+\phi_2^2)\right) \nonumber \\ &&\Theta \left(\eta_1^2+\phi_1^2-R^2 \right).
\eea
Let us change the variables to
\bea
\eta_1-\eta_2 &=& \lambda \cos \beta \, , \, \phi_1-\phi_2 = \lambda \sin \beta \nonumber \\
\eta_2 &=& \rho \cos \alpha
\, , \, \phi_2 = \rho \sin \alpha.
\eea
Doing so and taking care of the fact that $-\pi < \phi_1 < \pi$ we get 
\begin{multline}
\int \frac{\cos \left(\lambda \sin \beta\right)}{\cosh \left( \lambda \cos \beta \right)-\cos \left(\lambda \sin \beta\right)}\Theta \left [\rho^2+\lambda^2+2 \rho \lambda\cos(\alpha-\beta) -R^2 \right] \Theta \left [R^2-\rho^2 \right]\\ 
\Theta\left [ \pi-(\rho \sin \alpha+\lambda \sin \beta) \right ]\Theta \left[ \pi+(\rho \sin \alpha+\lambda \sin \beta) \right]  \rho d\rho \frac{d\alpha}{2\pi} \lambda d\lambda \frac{d\beta}{2\pi} 
\end{multline}
The step function implies 
\begin{equation}
\cos (\alpha-\beta) > \frac{R^2-\rho^2-\lambda^2}{2\rho \lambda}
\end{equation}
This is automatically satisfied if $(R^2-\rho^2-\lambda^2)/(2 \rho \lambda) <-1$. It is never satisfied if $(R^2-\rho^2-\lambda^2)/(2 \rho \lambda) >1$. Hence, the step function in question is only active when $-1 <(R^2-\rho^2-\lambda^2)/(2 \rho \lambda) < 1$. For a non-zero result we must then have $\rho+\lambda>R$.

Let us consider first the region of integration where the first step function is active. We have two integrals
\begin{multline}
 \int_0^R\lambda d\lambda\frac{\cos (\lambda \sin \beta)}{\cosh(\lambda \cos \beta)-\cos(\lambda \sin \beta)} \frac{d\beta}{2\pi} \int_{R-\lambda}^{R}\rho d\rho \frac{d\alpha}{2\pi} \Theta \left(\rho^2+\lambda^2+2\rho \lambda \cos(\alpha-\beta) -R^2 \right)\\
 + \int_R^{2 R}\lambda d\lambda\frac{\cos (\lambda \sin \beta)}{\cosh(\lambda \cos \beta)-\cos(\lambda \sin \beta)} \frac{d\beta}{2\pi} \int_{\lambda-R}^{R}\rho d\rho \frac{d\alpha}{2\pi} \Theta \left(\rho^2+\lambda^2+2\rho \lambda \cos(\alpha-\beta) -R^2 \right),
\end{multline}
 where we have not written the other step functions that are automatically 
satisifed since $\lambda$ and $\rho$ are small here. The first integral yields 
$\approx 0.61 R^2$ while the second one is $\approx 0.37 R^2$.

 When the cosine condition is automatically satisfied we have the following range of values 
\begin{equation}
R< \lambda <2R , \, 0<\rho<\lambda-R 
\end{equation}
and 
\begin{equation}
 \lambda> 2R, \, 0< \rho< R.
\end{equation}
The first integral is
\begin{equation}
\int_R^{2R} \lambda d\lambda\frac{\cos (\lambda \sin \beta)}{\cosh(\lambda \cos \beta)-\cos(\lambda \sin \beta)} \frac{d\beta}{2\pi} \int_0^{\lambda-R}\rho d\rho \frac{d\alpha}{2\pi}
\end{equation}
Note that for small $R$ and the above range of values of $\rho$ all other theta functions are also satisfied as e.g it is always true that $\pi > \rho\sin \alpha+\lambda \sin \beta$. One can evaluate the integral above easily and the result is $\approx 0.19 R^2$.

The remaining integral is 
\begin{multline}
\int_{2R}^{\infty} \lambda d\lambda\frac{\cos (\lambda \sin \beta)}{\cosh(\lambda \cos \beta)-\cos(\lambda \sin \beta)} \frac{d\beta}{2\pi} \int_0^{R}\rho d\rho \frac{d\alpha}{2\pi} \\ \Theta\left [ \pi-(\rho \sin \alpha+\lambda \sin \beta) \right ]\Theta \left[ \pi+(\rho \sin \alpha+\lambda \sin \beta) \right ]
\end{multline}
We note that we can split the $\lambda$ integral to go from $2 R$ to unity and then unity to infinity. The integral from $2R$ to unity produces the following result: 
$$-R^2 \ln 2R -0.12 R^2$$ which leaves to evaluate the integral 
\begin{multline}
\int_{1}^{\infty} \lambda d\lambda\frac{\cos (\lambda \sin \beta)}{\cosh(\lambda \cos \beta)-\cos(\lambda \sin \beta)} \frac{d\beta}{2\pi} \int_0^{R}\rho d\rho \frac{d\alpha}{2\pi} \\ \Theta\left [ \pi-(\rho \sin \alpha+\lambda \sin \beta) \right ]\Theta \left[ \pi+(\rho \sin \alpha+\lambda \sin \beta) \right ]
\end{multline}
This goes as $k R^2$ with $k$ a constant to be determined. To evaluate $k$ it proves simplest to first take the derivative with respect to$R^2$ which turns the theta function involving $\rho$ into a delta function. The $\rho$ integration is then trivial and the rest of the integral can be done numerically and yields $k=0.12$. We thus have that the overall result is $0.12 R^2$. Note that it is interesting that this cancels the $-0.12 R^2$ obtained previously. Hence the full result is 
\begin{equation}
I_{12}\approx 4\left[1.17R^2 -R^2 \ln 2R -0.12 R^2+0.12R^2 \right] + \ord(R^4) \approx 4 \left[1.17 R^2 -R^2 \ln 2R\right] + \ord(R^4).
\end{equation}
The above expansion well describes the full answer for $I_{12}$ up to values of $R\approx 0.6$. For larger values of the jet radius we use a full numerical evaluation of the coefficient $I_{12}$.

Next we consider the in-recoil dipole $I_{14}$. 
The result can be expressed in terms of the dipole $I_{12}$ and a rapidity dependent function $\kappa(\Delta y)$, which we evaluate numerically
\begin{equation}
\frac{\left(1+e^{\Delta y}\right)^2}{\left(1+\cosh \Delta y \right)^2} \left(1.17 R^2 -R^2 \ln 2 R\right)+\frac{(1+e^{\Delta y})}{1+\cosh \Delta y} \kappa (\Delta y) R^2.
\end{equation}
Finally, the dipole $I_{24}$ is easily obtained as 
\begin{equation}
I_{24}= I_{14}(-\Delta y).
\end{equation}

\section{Resummation formulae} \label{app:resum}
In this section we collect the explicit expressions of the function $f_i$ which build up the resummed results:
\begin{eqnarray} \label{f_func}
f_1(\lambda) &=& - \frac{1}{2 \pi \beta_0 \lambda} \left [ \left(1-2 \lambda \right ) 
\ln \left(1-2\lambda \right)-2 \left ( 1-\lambda \right ) \ln \left
  (1-\lambda \right ) \right ],  \nonumber\\
f_2(\lambda) &=& - \frac{K}{4 \pi^2 \beta_0^2} \left [2 \ln \left 
(1-\lambda \right ) - \ln \left (1-2 \lambda \right )\right ] 
\nonumber \\  &&-\frac{ \beta_1}{2 \pi \beta_0^3} \left [ \ln \left (1-2\lambda \right )-2 \ln 
\left (1-\lambda \right ) + \frac{1}{2} \ln^2 \left (1- 2 \lambda \right ) 
- \ln^2 \left (1-\lambda \right ) \right ], \nonumber\\
f_{{\rm coll}, q}(\lambda)&=&- \frac{3}{4 \pi \beta_0} \ln \left ( 1-\lambda \right ), \nonumber \\
f_{{\rm coll}, g}(\lambda)&=&- \frac{1}{C_A} \ln \left ( 1-\lambda \right ), \nonumber \\
f_{{\rm l.a.}}(\lambda)&=&\frac{1}{4 \pi \beta_0} \ln (1-2 \lambda).
\end{eqnarray}
 $\lambda = \beta_0 \alpha_s L, \; L = \ln \frac{R^2}{v}$ and $\alpha_s =\alpha_s\left(p_T R \right)$ is the $\overline{\rm MS}$ strong coupling, which at aimed accuracy we need to consider at two-loops:
\begin{equation}
  \label{eq:twoloop-as}
  \alpha_s(k_t^2) = 
  \frac{\alpha_s(\mu^2)}{1-\rho}\left[1-\alpha_s(\mu^2)\frac{\beta_1}{\beta_0}
    \frac{\ln(1-\rho)}{1-\rho}\right] \,,\qquad
  \rho = \alpha_s(\mu^2) \beta_0 \ln\frac{\mu^2}{k_t^2}\,,
\end{equation}
where the coefficients of the QCD $\beta$-function are defined as
 \begin{equation}
\beta_0 = \frac{11 C_A - 2 n_f }{12 \pi}, \; \beta_1 = \frac{17 C_A^2 - 5 C_A n_f -3 C_F n_f}{24 \pi^2}\,,
\end{equation}
and the constant $K$ is given by
\begin{equation}
K = C_A \left (\frac{67}{18}- \frac{\pi^2}{6} \right ) - \frac{5}{9} n_f\,.
\end{equation}

Finally, we report the coefficients $a,b, c$ obtained  by fitting the functional form
\begin{equation}
f_{ij}(t)=  \frac{1+(a_{ij} t)^2}{1+(b_{ij}t)^{c_{ij}} } t^2
\end{equation}
to the numerical dipole evolution for non-global logarithms. 
The results are reported in Tables~\ref{table1} and \ref{table2}. 
Note that the results for $I_{14}$ and $I_{24}$ depend on the rapidity separation between the jets and they have been computed for $|\Delta y|=2$.
\begin{table}[th]
\centering
\begin{tabular}[c]{c|c||c|c|c|}
$R$ & $I_{12}$ & $a$ & $b$ & $c$ \\
\hline 
$0.1$  & $0.11$ & $2.67\,C_A$ & $0.93\,C_A$ & $1.66$ \\
$0.2$  & $0.34$ & $2.27\,C_A$ & $1.08\,C_A$ & $1.66$ \\ 
$0.3$  & $0.62$ & $2.05\,C_A$ & $1.14\,C_A$ & $1.66$ \\
$0.4$  & $0.92$ & $1.99\,C_A$ & $1.31\,C_A$ & $1.66$ \\
$0.5$  & $1.22$ & $0.97\,C_A$ & $0.00\,C_A$ & $1.33$ \\
$0.6$  & $1.52$ & $0.85\,C_A$ & $0.00\,C_A$ & $1.33$ \\
$0.7$  & $1.80$ & $0.96\,C_A$ & $0.29\,C_A$ & $1.33$ \\ 
$0.8$  & $2.05$ & $1.12\,C_A$ & $0.56\,C_A$ & $1.33$ \\
$0.9$  & $2.28$ & $1.12\,C_A$ & $0.63\,C_A$ & $1.33$ \\
$1.0$  & $2.49$ & $1.24\,C_A$ & $0.80\,C_A$ & $1.33$ \\
$1.1$  & $2.67$ & $1.33\,C_A$ & $0.98\,C_A$ & $1.33$ \\ 
$1.2$  & $2.82$ & $1.45\,C_A$ & $1.15\,C_A$ & $1.33$ \\
\hline
\end{tabular}
\begin{tabular}{|c||c|c|c|}
 $I_{13}$& $a$         & $b$         & $c$ \\
\hline
 $3.289$   & $0.99\,C_A$ & $1.06\,C_A$ & $1.33$ \\
 $3.289$   & $0.99\,C_A$ & $1.06\,C_A$ & $1.33$ \\
 $3.289$   & $0.99\,C_A$ & $1.06\,C_A$ & $1.33$ \\
 $3.290$   & $0.94\,C_A$ & $0.99\,C_A$ & $1.33$ \\
 $3.290$   & $0.96\,C_A$ & $0.98\,C_A$ & $1.33$ \\
 $3.292$   & $0.86\,C_A$ & $0.87\,C_A$ & $1.33$ \\
 $3.293$   & $0.79\,C_A$ & $0.82\,C_A$ & $1.33$ \\
 $3.295$   & $0.79\,C_A$ & $0.81\,C_A$ & $1.33$ \\
 $3.299$   & $0.77\,C_A$ & $0.78\,C_A$ & $1.33$ \\
 $3.303$   & $0.86\,C_A$ & $0.85\,C_A$ & $1.33$ \\
 $3.310$   & $0.98\,C_A$ & $0.99\,C_A$ & $1.33$ \\
 $3.318$   & $0.91\,C_A$ & $0.92\,C_A$ & $1.33$ \\
\hline
\end{tabular}

\caption{Numerical results for the coefficients that parametrize the resummation of non-global logarithms.} \label{table1}
\end{table}
\begin{table}[th]
\centering
\begin{tabular}[c]{c|c||c|c|c|}
$R$ & $I_{14}$ & $a$ & $b$ & $c$ \\
\hline 
$0.1$  & $0.09$ & $2.74\,C_A$ & $0.86\,C_A$ & $1.66$ \\
$0.2$  & $0.28$ & $2.57\,C_A$ & $1.27\,C_A$ & $1.66$ \\ 
$0.3$  & $0.51$ & $2.27\,C_A$ & $1.31\,C_A$ & $1.66$ \\
$0.4$  & $0.77$ & $2.02\,C_A$ & $1.25\,C_A$ & $1.66$ \\
$0.5$  & $1.04$ & $1.03\,C_A$ & $0.00\,C_A$ & $1.33$ \\
$0.6$  & $1.31$ & $0.94\,C_A$ & $0.00\,C_A$ & $1.33$ \\
$0.7$  & $1.57$ & $0.93\,C_A$ & $0.14\,C_A$ & $1.33$ \\ 
$0.8$  & $1.81$ & $0.96\,C_A$ & $0.31\,C_A$ & $1.33$ \\
$0.9$  & $2.04$ & $0.86\,C_A$ & $0.21\,C_A$ & $1.33$ \\
$1.0$  & $2.25$ & $0.90\,C_A$ & $0.30\,C_A$ & $1.33$ \\
$1.1$  & $2.44$ & $1.18\,C_A$ & $0.72\,C_A$ & $1.33$ \\ 
$1.2$  & $2.61$ & $1.00\,C_A$ & $0.51\,C_A$ & $1.33$ \\
\hline
\end{tabular}
\centering
\begin{tabular}[c]{|c||c|c|c|c||}
 $I_{24}$ & $a$ & $b$ & $c$ \\
\hline 
   $0.00$ & $3.85\,C_A$ & $0.00\,C_A$ & $2.00$ \\
  $0.01$ & $3.28\,C_A$ & $0.31\,C_A$ & $2.00$ \\ 
   $0.02$ & $2.84\,C_A$ & $0.23\,C_A$ & $2.00$ \\
   $0.03$ & $2.89\,C_A$ & $0.59\,C_A$ & $1.66$ \\
 $0.05$ & $2.90\,C_A$ & $0.77\,C_A$ & $1.66$ \\
 $0.07$ & $2.76\,C_A$ & $0.81\,C_A$ & $1.66$ \\
 $0.10$ & $2.53\,C_A$ & $0.75\,C_A$ & $1.66$ \\ 
 $0.13$ & $2.45\,C_A$ & $0.78\,C_A$ & $1.66$ \\
 $0.16$ & $2.50\,C_A$ & $0.93\,C_A$ & $1.66$ \\
 $0.20$ & $2.45\,C_A$ & $0.99\,C_A$ & $1.66$ \\
 $0.24$ & $2.39\,C_A$ & $1.05\,C_A$ & $1.66$ \\ 
 $0.30$ & $2.56\,C_A$ & $1.30\,C_A$ & $1.66$ \\
\hline
\end{tabular}
\caption{Numerical results for the coefficients that parametrize the resummation of non-global logarithms. Note that the above results are valid in the case $|\Delta y|=2$.} \label{table2}
\end{table}


\begin{thebibliography}{99}

\baselineskip14pt
\bibitem{boost1}
 A.~Abdesselam {\it et al.},
  Eur.\ Phys.\ J.\ C {\bf 71} (2011) 1661
  [arXiv:1012.5412 [hep-ph]].
  
  \bibitem{boost2}
  A.~Altheimer{\it et al.},
  J.\ Phys.\ G G {\bf 39} (2012) 063001
  [arXiv:1201.0008 [hep-ph]].


 
\bibitem{Seymour}
  M.~H.~Seymour,
  Z.\ Phys.\  C {\bf 62} (1994) 127.

\bibitem{BDRS}
  J.~M.~Butterworth, A.~R.~Davison, M.~Rubin and G.~P.~Salam,
  Phys.\ Rev.\ Lett.\  {\bf 100} (2008) 242001
  [arXiv:0802.2470 [hep-ph]].
  
  \bibitem{atlas_HZ}
  G.~Aad {\it et al.}  [ATLAS Collaboration],
  ATL-PHYS-PUB-2009-088.
  
\bibitem{PSS}
   T.~Plehn, G.~P.~Salam and M.~Spannowsky,
  Phys.\ Rev.\ Lett.\  {\bf 104} (2010) 111801
  [arXiv:0910.5472 [hep-ph]].
  
  \bibitem{nlojet}
  Z.~Nagy,
  Phys.\ Rev.\  D {\bf 68} (2003) 094002
  [arXiv:hep-ph/0307268].

\bibitem{POWHEG}
P.Nason,
P.~Nason,
  JHEP {\bf 0411} (2004) 040
  [hep-ph/0409146].
S.~Frixione, P.~Nason and C.~Oleari,
  JHEP {\bf 0711} (2007) 070
  [arXiv:0709.2092 [hep-ph]].

\bibitem{MC@NLO}
 S.~Frixione and B.~R.~Webber,
  JHEP {\bf 0206} (2002) 029
  [hep-ph/0204244].
  
\bibitem{Dassalmag}
M.~Dasgupta, L.~Magnea and G.~P.~Salam,
  JHEP {\bf 0802} (2008) 055
  [arXiv:0712.3014 [hep-ph]].


\bibitem{ATLASjetshape}
  G.~Aad {\it et al.}  [ATLAS Collaboration],
  Phys.\ Rev.\ D {\bf 83} (2011) 052003
  [arXiv:1101.0070 [hep-ex]].
  
  \bibitem{ATLASjetmass}
  G.~Aad {\it et al.}  [ATLAS Collaboration],
  JHEP {\bf 1205} (2012) 128
  [arXiv:1203.4606 [hep-ex]].

    \bibitem{ATLASboosted}
 G.~Aad {\it et al.}  [ATLAS Collaboration],
  arXiv:1206.5369 [hep-ex].
  
  \bibitem{SchwaBec}
 T.~Becher and M.~D.~Schwartz,
  JHEP {\bf 0807} (2008) 034
  [arXiv:0803.0342 [hep-ph]].

\bibitem{CTTW}
  S.~Catani, L.~Trentadue, G.~Turnock and B.~R.~Webber,
  Nucl.\ Phys.\  B {\bf 407} (1993) 3.

\bibitem{BSZpheno}
  A.~Banfi, G.~P.~Salam and G.~Zanderighi,
  JHEP {\bf 1006} (2010) 038
  [arXiv:1001.4082 [hep-ph]].

\bibitem{EHLVW1}
  S.~D.~Ellis, A.~Hornig, C.~Lee, C.~K.~Vermilion and J.~R.~Walsh,
  arXiv:0912.0262 [hep-ph].
  
\bibitem{EHLVW2}
  S.~D.~Ellis, A.~Hornig, C.~Lee, C.~K.~Vermilion and J.~R.~Walsh,
  arXiv:1001.0014 [hep-ph].


\bibitem{yuan1}
  H.~-n.~Li, Z.~Li and C.~-P.~Yuan,
  Phys.\ Rev.\ Lett.\  {\bf 107} (2011) 152001
  [arXiv:1107.4535 [hep-ph]].

\bibitem{yuan2}
  H.~-n.~Li, Z.~Li and C.~-P.~Yuan,
  arXiv:1206.1344 [hep-ph].


\bibitem{DassalNG1}
  M.~Dasgupta and G.~P.~Salam,
  Phys.\ Lett.\  B {\bf 512} (2001) 323
  [arXiv:hep-ph/0104277].

  \bibitem{DassalNG2}
  M.~Dasgupta and G.~P.~Salam,
  JHEP {\bf 0203} (2002) 017
  [arXiv:hep-ph/0203009].

\bibitem{BDKM}
  A.~Banfi, M.~Dasgupta, K.~Khelifa-Kerfa and S.~Marzani,
  JHEP {\bf 1008} (2010) 064
  [arXiv:1004.3483 [hep-ph]].
  

\bibitem{KSZ1}
  R.~Kelley, M.~D.~Schwartz and H.~X.~Zhu,
  arXiv:1102.0561 [hep-ph].
  
  \bibitem{KSZ2}
  R.~Kelley, M.~D.~Schwartz, R.~M.~Schabinger and H.~X.~Zhu,
  arXiv:1112.3343 [hep-ph].
  
\bibitem{DasSalDIS}
  M.~Dasgupta and G.~P.~Salam,
  JHEP {\bf 0208} (2002) 032
  [arXiv:hep-ph/0208073].
\bibitem{antikt}
  M.~Cacciari, G.~P.~Salam and G.~Soyez,
  JHEP {\bf 0804} (2008) 063
  [arXiv:0802.1189 [hep-ph]].

\bibitem{kt1}
 S.~Catani, Y.~L.~Dokshitzer, M.~H.~Seymour and B.~R.~Webber,
  Nucl.\ Phys.\  B {\bf 406} (1993) 187.

  \bibitem{kt2}
  S.~D.~Ellis and D.~E.~Soper,
  Phys.\ Rev.\  D {\bf 48} (1993) 3160
  [arXiv:hep-ph/9305266].
  
 \bibitem{CAM}
  Y.~L.~Dokshitzer, G.~D.~Leder, S.~Moretti and B.~R.~Webber,
  JHEP {\bf 9708} (1997) 001
  [arXiv:hep-ph/9707323].

  \bibitem{CA}
  M.~Wobisch and T.~Wengler,
  arXiv:hep-ph/9907280.


 \bibitem{BanDas05}
  A.~Banfi and M.~Dasgupta,
  Phys.\ Lett.\  B {\bf 628} (2005) 49
  [arXiv:hep-ph/0508159].
  
  \bibitem{BanDasDel}
  Y.~Delenda, R.~Appleby, M.~Dasgupta and A.~Banfi,
  JHEP {\bf 0612} (2006) 044
  [arXiv:hep-ph/0610242].
  
  \bibitem{KKK}
  K.~Khelifa-Kerfa,
  JHEP {\bf 1202} (2012) 072
  [arXiv:1111.2016 [hep-ph]].
  
  \bibitem{KWZ1}
  R.~Kelley, J.~R.~Walsh and S.~Zuberi,
  arXiv:1202.2361 [hep-ph].
  
  \bibitem{KWZ2}
  R.~Kelley, J.~R.~Walsh and S.~Zuberi,
  arXiv:1203.2923 [hep-ph].
  

  
\bibitem{Dassalrev}
 M.~Dasgupta and G.~P.~Salam,
  J.\ Phys.\ G G {\bf 30} (2004) R143
  [hep-ph/0312283].


\bibitem{BSZcaesar}
  A.~Banfi, G.~P.~Salam and G.~Zanderighi,
  JHEP {\bf 0503} (2005) 073
  [hep-ph/0407286].
  

    
    
      \bibitem{KOS}
  N.~Kidonakis, G.~Oderda and G.~F.~Sterman,
  Nucl.\ Phys.\ B {\bf 531} (1998) 365
  [hep-ph/9803241].

\bibitem{BMS}
  A.~Banfi, G.~Marchesini and G.~Smye,
  JHEP {\bf 0208} (2002) 006
  [arXiv:hep-ph/0206076].


\bibitem{QCDESW}
 R.K.~Ellis, W.J.~Stirling and B.R.~Webber,
Cambridge Monographs on Particle Physics, Nuclear Physics and Cosmology

\bibitem{Catani:1990rr}
  S.~Catani, B.~R.~Webber and G.~Marchesini,
  Nucl.\ Phys.\ B {\bf 349} (1991) 635.

\bibitem{FKM}
  J.~Forshaw, J.~Keates and S.~Marzani,
  JHEP {\bf 0907} (2009) 023
  [arXiv:0905.1350 [hep-ph]].

\bibitem{CatGraz2gluon}
 S.~Catani and M.~Grazzini,
  Nucl.\ Phys.\ B {\bf 570} (2000) 287
  [hep-ph/9908523].
  
\bibitem{BMDZ3jet}
  A.~Banfi, G.~Marchesini, Y.~L.~Dokshitzer and G.~Zanderighi,
  JHEP {\bf 0007} (2000) 002
  [hep-ph/0004027].

\bibitem{milan}
  Y.~L.~Dokshitzer, A.~Lucenti, G.~Marchesini and G.~P.~Salam,
  Nucl.\ Phys.\ B {\bf 511} (1998) 396
   [Erratum-ibid.\ B {\bf 593} (2001) 729]
  [hep-ph/9707532].

 
  \bibitem{HLWZinout}
  A.~Hornig, C.~Lee, J.~R.~Walsh and S.~Zuberi,
  JHEP {\bf 1201} (2012) 149
  [arXiv:1110.0004 [hep-ph]].


\bibitem{FKS1}
  J.~R.~Forshaw, A.~Kyrieleis and M.~H.~Seymour,
  JHEP {\bf 0608} (2006) 059
  [hep-ph/0604094].

\bibitem{FKS2}
  J.~R.~Forshaw, A.~Kyrieleis and M.~H.~Seymour,
  JHEP {\bf 0809} (2008) 128
  [arXiv:0808.1269 [hep-ph]].

\bibitem{DFMS}
 R.~M.~Duran Delgado, J.~R.~Forshaw, S.~Marzani and M.~H.~Seymour,
  JHEP {\bf 1108} (2011) 157
  [arXiv:1107.2084 [hep-ph]].
  
\bibitem{comix} 
  T.~Gleisberg and S.~Hoeche,
  JHEP {\bf 0812}, 039 (2008)
  [arXiv:0808.3674 [hep-ph]].

\bibitem{sherpa}
  T.~Gleisberg, S.~Hoeche, F.~Krauss, A.~Schalicke, S.~Schumann and J.~-C.~Winter,
  JHEP {\bf 0402}, 056 (2004);
   T.~Gleisberg, S.~.Hoeche, F.~Krauss, M.~Schonherr, S.~Schumann, F.~Siegert and J.~Winter,
  JHEP {\bf 0902}, 007 (2009).


 \bibitem{cteq6m}
  J.~Pumplin, D.~R.~Stump, J.~Huston, H.~L.~Lai, P.~M.~Nadolsky, W.~K.~Tung,
  JHEP {\bf 0207 } (2002)  012.
  [arXiv:hep-ph/0201195 [hep-ph]]. 


 \bibitem{mcfm}
  J.~M.~Campbell and R.~K.~Ellis,
  Phys.\ Rev.\  D {\bf 65} (2002) 113007
  [arXiv:hep-ph/0202176].

\bibitem{giantKfact}
  M.~Rubin, G.~P.~Salam and S.~Sapeta,
  JHEP {\bf 1009} (2010) 084
  [arXiv:1006.2144 [hep-ph]].

\bibitem{pythia}
  T.~Sjostrand, S.~Mrenna and P.~Z.~Skands,
  Comput.\ Phys.\ Commun.\  {\bf 178}, 852 (2008).

\bibitem{herwig}
M.~Bahr, S.~Gieseke, M.~A.~Gigg, D.~Grellscheid, K.~Hamilton, O.~Latunde-Dada, S.~Platzer and P.~Richardson {\it et al.},
  Eur.\ Phys.\ J.\ C {\bf 58}, 639 (2008);
S.~Gieseke, D.~Grellscheid, K.~Hamilton, A.~Papaefstathiou, S.~Platzer, P.~Richardson, C.~A.~Rohr and P.~Ruzicka {\it et al.},
  arXiv:1102.1672 [hep-ph].

\bibitem{DokWeb97}
  Y.~L.~Dokshitzer and B.~R.~Webber,
  Phys.\ Lett.\ B {\bf 404} (1997) 321
  [hep-ph/9704298].

\end{thebibliography}
\end{document}